\documentclass[12pt,preprint]{aastex}
\begin{document}
\parindent=1.0cm

\title{The Evolved Stellar Content of NGC 147, NGC 185, and NGC 205}

\author{T. J. Davidge \altaffilmark{1} \altaffilmark{2}}

\affil{Canadian Gemini Office, Herzberg Institute of Astrophysics,
\\National Research Council of Canada, 5071 West Saanich Road,
\\Victoria, B.C. Canada V9E 2E7}

\email{tim.davidge@nrc.ca}

\altaffiltext{1}{Visiting Astronomer, Canada-France-Hawaii Telescope,
which is operated by the National Research Council of Canada, the Centre National de le
Recherche Scientifique, and the University of Hawaii.} 

\altaffiltext{2}{This publication 
makes use of data products from the Two Micron Sky Survey, which is a joint project of the
University of Massachusetts and the Infrared Processing and Analysis Center/California 
Institute of Technology, funded by NASA and the NSF.}

\begin{abstract}

	Broad-band near-infrared images are used to probe the 
photometric properties of the brightest asymptotic giant branch (AGB) stars within 2 
arcminutes of the centers of the dwarf elliptical galaxies NGC 147, NGC 185, and NGC 205. 
Sequences originating from oxygen-rich M giants and C stars are clearly 
distinguished on the $(K, H-K)$ and $(K, J-K)$ color-magnitude diagrams (CMDs). 
Based on the peak brightness and color of the M giant sequences, ages 
of 1 Gyr and 0.1 Gyr are predicted for the most recent significant star forming events in 
NGC 185 and NGC 205, respectively. The bolometric luminosity function (LF) of M giants in 
NGC 205 is flatter than in NGC 185, in agreement with studies at wavelengths shortward 
of $1\mu$m. The most luminous AGB stars in NGC 147 are 
well mixed with fainter stars throughout the area surveyed in this galaxy, 
and the peak brightness of the M giant sequence indicates that the most recent significant 
star-forming activity occured $\sim 3$ Gyr in the past.

	The C star contents of the three galaxies are compared in two ways. First, 
the notion of a specific frequency measurement for C stars, in which C star 
counts per unit magnitude are normalised to a common integrated $K-$band brightness using 
published surface photometry, is introduced. The specific frequency of C stars 
outside of the areas of most recent star formation is found to agree in all three 
galaxies. Second, comparisons are made using the integrated brightness from C stars, which 
is normalised to the integrated light from M giants and the integrated light from all 
stars. The fractional contribution made by C stars to the total AGB light in the $K-$band 
is found to be highest in NGC 147 and lowest in the central regions of NGC 
205, which is qualitatively consistent with model predictions. 
The fractional contribution that C stars make to the total $K-$band light 
is found to be constant both within and between galaxies, with C stars contributing 2\% 
of the total $K-$band light. It is concluded that, 
when averaged over timescales of a few Gyr, these galaxies have turned 
similar fractions of gas and dust, normalised according to total galaxy mass, into stars. 
It is argued that the material for star formation likely originated 
in well-regulated reservoirs, and it is demonstrated that the mass of gas needed to 
fuel star formation during intermediate epochs could have been supplied by winds from 
evolved stars. Finally, multi-epoch data are used to investigate the incidence of long 
period variables (LPVs) in NGC 185 and NGC 205. While tight constraints can not 
be placed on the LPV content of NGC 205, roughly 70\% 
of the stars with M$_K$ between --7.5 and --8.0 in the central regions of NGC 185 appear to 
be LPVs with amplitudes similar to those of LPVs in the LMC. 

\end{abstract}


\section{INTRODUCTION}

	Spiral galaxies are accompanied by satellites that have a range of 
structural characteristics. The brightest companions of some spiral galaxies tend to be 
disks, while in other cases the brightest companions are spheroids. This 
diversity in companion properties is clearly evident among the closest 
large spiral galaxies. The brightest satellites of the Milky-Way are the 
disk-dominated LMC and SMC, both of which show moderate levels of star formation at the 
present day. For comparison, the brightest companions of the next nearest large spiral 
galaxy, M31, tend to be gas-poor spheroids, including the 
compact elliptical galaxy M32 and the dwarf ellipticals (dEs) NGC 147, NGC 185, and NGC 
205. \footnote[3]{While M31 also has gas-rich dwarf irregular companions (e.g. Evans et al. 
2000; Courteau \& van den Bergh 1999; Mateo 1998), only two (IC 1613 and IC 10) have 
integrated brightnesses that are comparable to those of M32 and the three dEs. These dwarf 
irregular galaxies are at very large galacto-centric distances ($\geq 250$ kpc), and IC 1613 
may not even be bound to M31 (e.g. Courteau \& van den Bergh 1999; Mateo 1998).} The next 
closest large spiral galaxy - M81 - is accompanied by a mixture of disk 
and spheroid galaxies that have a broad range of star-forming properties, some of 
which are experiencing elevated levels of star formation due to tidal interactions 
(e.g. Yun, Ho, \& Lo 1994). Despite the differences among the most easily studied high 
surface brightness satellites, M81, M31, and the Milky-Way are all accompanied by 
an entourage of dwarf spheroidal (dSph) galaxies that show broadly similar characteristics 
(Karachentev et al. 2001).

	Initial conditions, such as the fractional dark matter content, could affect 
morphology (e.g. Ferrara \& Tolstoy 2000; Carraro et al. 2001), and so are a 
possible cause of the range in structural properties seen among the satellite systems. 
Studies of dE's in the Virgo cluster (Geha, Guhathakurta, \& van der Marel 
2002) and the Local Group (e.g. Held et al. 1992) find M/L ratios that are 
suggestive of no significant dark matter contribution, at least within the visible 
portions of these systems. For comparison, Local Group dSphs appear to have high M/L ratios 
that are suggestive of a large dark matter halo (e.g. Irwin \& Hatzidimitriou 1995; 
Kleyna et al. 2002). As for dwarf irregular galaxies, Cote, Carignan, \& Freeman (2000) find 
that the M/L ratios of these objects in the Sculptor and 
Centaurus groups span a range of values, implying a dispersion in dark matter contents. 

	Environment can also influence morphology; for example, it has been suggested that 
galaxy-galaxy encounters in dense environments and tidal interactions with larger systems 
may transform disk systems into spheroids (e.g. Moore, Lake, \& Katz 
1998; Mayer et al. 2001). In extreme cases tidal interactions will completely 
disrupt some satellites, with only the dense central regions surviving to the present day 
and the stars that once belonged to the disrupted system being distributed along debris 
trails (e.g. Bekki \& Chiba 2004; Mizutani, Chiba, \& Sakamoto 2003; Bekki \& Freeman 
2003). It has been suggested that M32 may have evolved in such as 
manner, starting as a disk galaxy that was transformed by interactions with 
M31; Graham (2002) argues that signatures of the disk remain to the present day. 

	The stellar contents of satellites provide a fossil record that can be mined to
reveal their histories and gauge the impact that any interactions with larger 
companions have had on their evolution. NGC 147, NGC 185, and NGC 205 are 
important targets as they are similar to more distant 
dE galaxies (e.g. Zinnecker \& Cannon 1986), and have 
a wide range of projected distances from M31, raising the possiblity that 
the effects of environment might be de-coupled from any intrinsic properties of the 
galaxies. For example, a correlation between the number of stars that formed during the 
past few Gyr and distance from M31 might indicate that interactions with M31 played a key 
role in spurring star formation in these galaxies, and might even have provided the 
gas and dust necessary to stoke star formation. On the other hand, a discovery of 
similar star forming histories, despite the differences in environment, would favour 
the notion that the evolution of these galaxies is dominated by intrinsic properties.

	Given their importance as the nearest dEs, it is not surprising that the three M31 
dE companions have been the target of a number of investigations. NGC 147 has been 
found to contain a modest population of relatively bright AGB stars, that are the brightest 
members of a population that accounts for only a small fraction of the total 
$V-$band light (Davidge 1994). The absence of main sequence turn-off stars with 
M$_V < -1$ indicates that the most recent large scale star-forming activity occured 
at least 1 Gyr in the past (Han et al. 1997). NGC 147 contains C stars (e.g. Nowotny et al. 
2003). Battinelli \& Demers (2004a) find that the C stars are uniformly mixed throughout the 
main body of the galaxy, and may be slightly fainter on average in the $I-$band 
than their counterparts in other Local Group galaxies. There is a large population of 
older stars, and the color of the RGB indicates that [Fe/H] is between --0.9 
and --1.2, with a dispersion about the mean value and possible radial 
gradients in both metallicity and age (Mould, Kristian, \& Da Costa 1983; Davidge 1994; 
Han et al. 1997). A spread in metallicity indicates that NGC 147 has 
experienced chemical enrichment. Nevertheless, HI has yet to be detected in NGC 147 
(Young \& Lo 1997), and the upper limit computed for the ISM mass is markedly lower than 
is expected if material ejected from evolved stars has been retained by the galaxy, 
suggesting that the ISM has been depleted (Sage, Welch, \& Mitchell 1998). 

	NGC 185 contains stars that span a broader range of ages than in NGC 147 (Lee, 
Freedman, \& Madore 1993; Martinez-Delgado \& Aparicio 1998; Martinez-Delgado, Aparicio, 
\& Gallart 1999). The youngest stars have an age near 400 Myr 
(Butler \& Martinez-Delgado 2005) and are concentrated 
in the central $150 \times 90$ parsec$^2$; outside of this area there are stars 
with ages of at least 1 Gyr, which in turn are more centrally concentrated than even older 
populations (Martinez-Delgado et al. 1999). Kang et al. (2005) surveyed the AGB content of 
the central regions of NGC 185 at near-infrared wavelengths, and found 73 C stars. 
The mean M$_I$ of C stars in NGC 185 is comparable to that in NGC 147, although the C stars 
in NGC 185 are more centrally concentrated than in NGC 147 
(Battinelli \& Demers 2004b; Nowotny et al. 2003). The mean 
metallicity of the RGB is [Fe/H] = --1.2, with a dispersion of $\pm 0.3$ dex (Lee et al. 
1993). The ISM in NGC 185 is concentrated near the present-day star-forming region, 
although the atomic and molecular components are not spatially coincident (Welch, 
Mitchell, \& Yi 1996; Young 2001). The morphology and kinematic properties of the ISM 
are consistent with it originating from stars internal to NGC 185 
(Young \& Lo 1997; Martinez-Delgado et al. 1999). 

	NGC 205 is the brightest of the three dE companions of M31. 
There is a centrally concentrated population of 
young blue stars that have ages in the range $50 - 100$ Myr (Cappellari et al. 1999), 
and a solar metallicity AGB component that is uniformly mixed throughout 
the central 1 arcmin (Davidge 1992). Lee (1996) concludes that the AGB-tip in 
NGC 205 is $\sim 0.7$ mag brighter in $I$ than in NGC 185, and that 
the AGB luminosity functions (LFs) of these galaxies also differ, in the sense that the LF 
of NGC 205 is the flatter of the two. However, 
the C stars in NGC 205 have an average M$_I$ that is similar to what is seen 
in NGC 147 and NGC 185 (Demers, Battinelli, \& Letarte 2003). The color of the 
RGB suggests that [Fe/H] = --0.9 with a dispersion of $\pm 0.5$ dex (Mould, Kristian, \& 
Da Costa 1984). As with NGC 147, the total ISM mass of NGC 205 is less than expected if 
the mass lost from stars has been retained (Welch, Sage, \& Mitchell 1998). 

	NGC 205 is the closest dE to M31; hence, it might 
be anticipated that M31 has had a greater impact on the evolution of NGC 205 than on the 
evolution of either NGC 147 or NGC 185. In fact, the orbits of NGC 205, M32, and M31 
indicate that there may have been past interactions (Cepa \& Beckman 1988), and
there are a number of other clues that support this notion. 
First, the star-forming history of NGC 205 inferred from AGB stars is correlated with 
the orbit about M31, in that the AGB content shows evidence for multiple 
episodes of star formation during the past Gyr, with the most recent occuring within 
the past 0.1 Gyr (Davidge 2003). Second, the gas and stars in NGC 205 
form distinct dynamical systems (Young \& Lo 1997; Welch et al. 1998), as 
might be expected if the molecular clouds have been tidally affected by interactions 
with M31. Third, the isophotal properties of NGC 205 and the presence of stellar 
debris trails in the vicinity of M31 are indicative of interactions 
(e.g. Ibata et al. 2001, Choi, Guhathakurta, \& Johnston 2002, and McConnachie et al. 
2004). Finally, there is an HI cloud 25 arcmin southwest of NGC 205 that 
may contain material that has been stripped from the galaxy (Thilker et al. 2004). 

	The majority of stellar content studies of NGC 147, NGC 185, and NGC 205 
have used data that sample the visible and red portions 
of the spectrum. This wavelength region is of critical importance for probing young, 
blue stars and the hotter red giants that belong to moderately young and very 
metal-poor old populations. However, the visible portion of the spectrum is not ideal for 
studies of the reddest stars, such as those evolving near the AGB-tip, as the 
brightnesses of these objects at visible wavelengths are affected by line blanketing. 
This causes the AGB to form a horizontal or even descending sequence 
on CMDs constructed from visible wavelength data, 
thereby complicating efforts to detect stars near the AGB-tip. 
Not only does the observed brightness of AGB-tip stars
at visible wavelengths not reflect the true luminosity of these objects, but the 
contrast between AGB-tip stars and bluer objects is diminished, such that AGB-tip 
stars may occur in a brightness regime where confusion with more numerous objects, such 
as RGB stars, becomes an issue.

	The problems caused by line blanketing are much reduced in the 
near-infrared, where the AGB forms a near-vertical sequence on CMDs (e.g. Davidge 2003).
The contrast with respect to bluer stars is enhanced, and the brightness of 
the AGB-tip is easier to measure than at visible wavelengths. Finally, C stars can also 
be selected using broad-band near-infrared colors (Davidge 2003, Hughes \& Wood 1990). 
Near-infrared color is a criterion that differs from the narrow-band filter
technique that is traditionally used to identify C stars in the red part of the spectrum 
(e.g. Richer, Crabtree, \& Pritchet 1984), and so a 
comparison of C star properties based on near-infrared 
color and narrow-band imaging is of interest. 

	In the present study, $JHK$ images are used to investigate the bright AGB 
contents of NGC 147, NGC 185, and NGC 205. The data were 
obtained during the same night and with the same instrument, and thus are well-suited for 
comparing the relative stellar contents of these systems. The 
core data for NGC 185 and NGC 205 are supplemented with $K-$band 
observations obtained with the same instrument during a previous observing run, which 
are used to investigate the frequency and amplitude distribution of variable stars.

	The distances and reddenings adopted in this study are listed in Table 1. 
The distances are based on the $I-$band RGB-tip brightnesses measured by McConnachie et 
al. (2005), while the reddenings are from the maps constructed by Schlegel, Finkbeiner, \& 
Davis (1998). The distances from the center of M31 in kpc, R$_{M31}$, 
assuming the RGB-tip distance for M31 computed 
by McConnachie et al. (2005), are listed in the last column of Table 1. 
Of the three dEs, NGC 185 is the most distant from M31. 

	The paper is structured as follows. Details of the observations and the 
procedures used to process the data are presented in \S 2, while 
the photometric measurements are discussed in \S 3. The general 
characteristics of the color-magnitude diagrams (CMDs) and LFs are discussed in \S 4, 
while galaxy-to-galaxy comparisons of the properties of M giants and C stars are 
the subject of \S 5. The properties of LPVs in NGC 185 and NGC 205 are examined in 
\S 6. A discussion and summary of the results follows in \S 7.

\section{OBSERVATIONS \& REDUCTIONS}

	Near-infrared images of the central regions of NGC 147, NGC 185, and NGC 205 
were recorded with the CFHTIR imager on the night of UT November 21/22 2002. 
The CFHTIR detector is a $1024 \times 1024$ HgCdTe 
array, with an image scale of 0.21 arcsec pixel$^{-1}$ when 
mounted at the Cassegrain focus of the 3.6 metre Canada-France-Hawaii Telescope. 
The data were recorded through $J, H,$ and $K'$ filters 
as a series of relatively short (20 -- 60 sec) exposures 
to prevent saturating the background sky and the brightest stars in each field.
The total integration time is 12 minutes filter$^{-1}$ galaxy$^{-1}$. 

	A separate set of $K'$ observations were recorded of NGC 185 and NGC 205 
with the CFHTIR during an observing run in June 2001, and these data are used in \S 6 to 
investigate the variability of bright AGB stars in these galaxies. The 
NGC 205 observations were discussed previously by Davidge (2003), and a complete 
description of those data can be found in that paper. The NGC 185 data 
were recorded on the night of UT June 5/6 2001. The total integration time is 8 
minutes, and thin clouds were present when the data were recorded. 

	Data needed to construct calibration frames that can correct for 
artifacts introduced by the telescope, the instrument, and the sky were recorded on each 
night of the 2001 and 2002 observing runs. Flat field images of 
the dome were obtained at the beginning of each night, while 
the dark current was monitored at the end of each night. The flat-field pattern and dark 
count rate did not vary during each run, and so the nightly calibration exposures 
were combined to create master flat-field and dark frames. Areas of the sky 
with low stellar density were also observed at various times throughout each night, and 
these data were used to construct calibration images that monitor interference fringe 
patterns and the thermal emission from warm objects along the optical path.

	The data were reduced using a standard pipeline for near-infrared images. 
The basic steps in the processing sequence are: (1) the subtraction of a dark frame, 
(2) the division by a flat-field frame, (3) the subtraction of the DC sky level from each 
image, and (4) the subtraction of the appropriate interference fringe $+$ thermal emission 
calibration frame. The processed images for each field $+$ 
filter combination were aligned to correct for offsets introduced during acquisition, and 
then median combined. The final step was to remove those portions of the combined images 
that do not have the full exposure time. The final $J$ images of the galaxies 
are shown in Figures 1, 2, and 3. Single stars in the final images have FWHMs in the 
range 0.7 -- 0.9 arcsec, depending on the galaxy.

\section{PHOTOMETRIC MEASUREMENTS AND ARTIFICIAL STAR EXPERIMENTS}

	Stellar brightnesses were measured with the point spread function (PSF) fitting 
routine ALLSTAR (Stetson \& Harris 1988). The co-ordinates, preliminary 
brightnesses, and PSFs that are used by ALLSTAR were obtained from routines in the 
DAOPHOT (Stetson 1987) photometry package. The PSFs for each image were constructed from at 
least 40 stars.

	Standard stars from Hawarden et al. (2001) were observed during both the June 
2001 and November 2002 runs, and these were used to define the photometric calibration. 
The consistency of the calibration was checked by comparing the brightnesses of stars 
with $K < 14.5$ with measurements in the 2MASS Point Source Catalogue (Cutri et al. 2003). 
Sources within 1 arcmin of the centers of NGC 185 and NGC 205 
were avoided when making these comparisons. The mean 
differences, in the sense CFHTIR -- 2MASS are $\Delta K = 0.00 \pm 0.14, \Delta(H-K) = 
-0.03 \pm 0.06$, and $\Delta(J-K) = -0.01 \pm 0.05$, where the quoted uncertainties 
are the standard deviations about the mean. The standard deviation is an upper limit to 
the actual dispersion due to photometric errors as some of the stars used in this 
comparison may be photometric variables. Nevertheless, it is clear that the photometric 
calibration is reliable to within a few hundredths of a magnitude.

	Artificial star experiments were run to investigate sample completeness 
and the random uncertainties in the photometry due to crowding and photon 
noise. Both the degree of completeness and the errors in the photometry depend on the 
projected stellar density, in the sense that incompleteness sets in at progressively 
brighter levels, while the uncertainties in the photometry at a given magnitude become 
larger, as stellar density increases. To account for these effects in a basic way, 
the field imaged for each galaxy was divided into two equal area regions centered 
on the galaxy nucleus; the photometric properties of stars in the area that includes 
the center of each galaxy (the `inner' region) are considered separately from 
those that are outside of this area (the `outer' region). 
This division also permits gradients in stellar content to be investigated (\S 5).

	The artificial stars were assigned colors that are representative 
of bright M giants and C stars, and were added in 0.5 magnitude increments 
in $K$. Based on the CMDs that are discussed in \S 4, 
M giants were assumed to have $H-K = 0.3$ and $J-K = 1.1$ 
while C stars were assumed to have $H-K = 0.9$ and $J-K = 1.8$. 
A total of sixty artificial stars of each type, evenly split between the inner and outer 
regions, were added to the images of each galaxy at each $K$ magnitude.

	The standard deviations in the differences between the 
measured and actual values of $K$, $H-K$ and $J-K$ for stars with $K = 17$, which is a 
brightness at which both M giants and C stars are seen in all three 
galaxies, are listed in Tables 2 (M giants) and 3 (C stars).
The entries in Tables 2 and 3 indicate that the photometric errors predicted for the 
inner and outer regions of all three galaxies differ by only modest amounts. Moreover, the 
uncertainties in the NGC 205 measurements are larger than those for either NGC 147 and NGC 
185, due to the higher stellar density in this galaxy. This complicates efforts to identify 
LPVs in NGC 205 (\S 6). 

	The relative uncertainties in brightnesses and colors in Tables 2 and 
3 warrant discussion. The uncertainties in the colors of M giants in Table 2 are 
smaller than the uncertainties in the $K$ brightnesses. This is because the factors that 
affect brightness measurements, such as crowding, are to some extent 
panchromatic, and so at least partly cancel out when colors are 
computed. Indeed, the results in Table 2 illustrate why artificial 
star experiments should be run with stars having realistic colors; it is not 
correct to compute the uncertainties in colors simply by adding the uncertainties in 
the brightness measurements in quadrature, as systematic effects are not considered. 

	The situation appears to be different in Table 3, where the uncertainties in the 
colors of C stars tend to be larger than those in the $K$ measurements. However, 
the uncertainties in the colors of C stars are larger than the uncertainties 
in the colors of M giants because the former have redder colors, which makes them 
fainter in $H$ and $J$ at a given $K$. As a result, the uncertainties in the $J$ and 
$H$ measurements are larger than those in $K$, and so play a larger role in determining 
the uncertainties in the colors. Nevertheless, the uncertainties in the colors are 
still smaller than what would be computed simply by adding the uncertainties in the 
individual brightnesses in quadrature. 

	The artificial star experiments can also be used to assess the effects of 
blending, in which two or more stars fall within the same angular resolution element, 
and so appear as a single object. Given that the CFHTIR data 
sample the crowded central regions of these galaxies then all stars are likely blends, 
although in the vast majority of cases the bright stars that are the topic of this 
paper are blended with an object that is much fainter, so that the photometric 
measurements are essentially those of the brighter star. However, in the 
rare event that two AGB stars with comparable brightness fall in the same resolution 
element then the result will appear as a single source that is --2.5 log(2) $= 0.75$ 
magnitude brighter than the progenitors.

	The histogram distribution of the difference between the actual and 
recovered $K-$band brightnesses of artificial stars representing 
M giants with $K = 17$, $\Delta K$, is shown in Figure 4 for each galaxy. Obvious blends 
have positive values of $\Delta K$, and so 
skew the distributions if they occur in large 
numbers. It is evident from Figure 4 that the $\Delta K$ distributions 
in the outer regions are symmetric about $\Delta K = 0$ with no 
significant tail to positive values, suggesting that blending does not have a major 
impact on the photometric properties of stars with $K = 17$ 
in these areas of the galaxies. While there are a handful of stars with 
$\Delta K \geq 0.5$ in the inner regions, these account for only a modest fraction of the 
total number of objects, and the main body of data is still centered 
near $\Delta$K $\sim 0$. The results shown in Figure 4 indicate that blending does not 
have a major impact on the photometric properties of stars
with $K = 17$ in these galaxies. The effects of blending will 
be even smaller when $K < 17$, but greater when $K > 17$.

\section{COLOR-MAGNITUDE DIAGRAMS, LUMINOSITY FUNCTIONS, AND THE SELECTION OF C STARS}

	The $(H, J-H)$, $(K, H-K)$, and $(K, J-K)$ CMDs of the galaxies are shown in 
Figures 5, 6, and 7. The dominant feature in these CMDs is a plume that contains 
luminous oxygen-rich M giants that are evolving on the 
AGB. The AGB plume is almost vertical in the $(H, J-H)$ CMDs, 
and the sequence in the NGC 205 $(H, J-H)$ CMD is clearly broader than in the other 
galaxies. This is consistent with the artificial star experiments, which predict larger 
uncertainties in the photometric measurements of stars in 
NGC 205 at a given brightness (\S 3). The $(K, H-K)$ 
and $(K, J-K)$ CMDs show more structure than the $(H, J-H)$ CMDs, in the sense that 
the AGB broadens at the bright end, with a spray of stars extending to red colors. 

	The spray of bright objects to the right of the M giant sequence in the CMDs 
is due to C stars. In order to isolate a clean sample of C stars it is necessary to 
determine the color at which C stars dominate over M giants. 
Hughes \& Wood (1991) investigated the near-infrared photometric properties of 
LPVs in the LMC, and concluded that stars with $J-K > 1.6$ and $H-K > 0.6$ are C stars. 
However, the majority of objects in the Hughes \& Wood (1990) study have $J-K < 1.4$ and 
$J-K > 2.0$, and the number of objects with $J-K \sim 1.6$ is modest. In fact, significant 
numbers of spectrographically confirmed C stars in the LMC and SMC have 
$J-K$ colors as small as 0.6 (e.g. Demers, Dallaire, \& Battinelli 
2002). While the identification of such warm C stars from broad-band photometric data 
alone is problematic, Demers et al. (2002) find that the majority of C stars identified 
from narrow-band imaging and red broad-band photometry have $(J-K)_0 > 1.4$, and 
advocate this as a criterion for near-infrared C star surveys. This color criterion is 
supported by other studies. Nikolaev \& Weinberg (2000) identify a tongue of objects with 
$(J-K)_0 > 1.4$ that departs from the M giant sequence 
on the $(K, J-K)$ CMD of the LMC that is associated with C stars. Cioni \& Habing (2005) 
identify a similar feature in the $(K, J-K)$ CMD of NGC 6822, which is also 
associated with a plateau in the $J-K$ color distribution when $(J-K)_0 > 1.4$. 
Based on these results, it is assumed in the present study that stars with $(J-K)_0 
> 1.4$ are C stars. The near-infrared two-color diagram for LMC LPVs constructed by 
Hughes \& Wood (1990) indicates that this corresponds to $(H-K)_0 = 0.45$.

	A brightness limit for C stars can also be set. The C star features seen in 
the $(K, J-K)$ CMDs of the LMC (Nikolaev \& Weinberg 2000) and NGC 6822 (Cioni \& Habing 
2005) have M$_K > -7.25$. Consequently, M$_K = -7.25$ is adopted as the faint limit for 
C stars in the present study.

	The color and brightness boundaries for identifying C stars are marked in 
Figures 6 and 7. The color boundary tracks the red edge of the M giant sequence in both the 
$(K, H-K)$ and $(K, J-K)$ CMDs, while a number of C stars are seen to the right of 
this boundary. It is worth noting that a C star sequence can be seen in 
the $(H, J-H)$ CMDs, although the C star sequence 
plunges rapidly downward towards red colors, and so is not as 
distinct as the C star sequences in the $(K, H-K)$ and $(K, J-K)$ CMDs. 
It is because the C star and M giant sequences are clearly separated on the $(K, H-K)$ and 
$(K, J-K)$ CMDs that these CMDs were adopted as the basis for the detailed analysis of 
AGB properties in the remainder of the paper.

	The brightest stars in each CMD belong to the 
foreground disk, and these form a loose sequence with relatively blue colors. 
NGC 147 and NGC 185 are viewed at lower Galactic latitudes than NGC 205, and so it is not 
surprising that the CMDs of these galaxies contain a larger number of foreground stars than 
in NGC 205. In any event, contamination from foreground stars is likely not a major issue 
when investigating the bright AGB population of these galaxies. Indeed, the foreground star 
sequence tends to be bluer than the brightest M giants in all three galaxies, although 
there is some overlap, and the foreground and AGB sequences only merge near the faint end 
of the CMDs, where the errors in the photometry are large. As for 
the region of the CMDs containing C stars, there 
appears to be no obvious contamination from foreground stars.

	While the number of foreground stars is modest, they may still skew efforts 
to study the brightest portions of the AGB. 
To assess the peak brightness of objects free of foreground star contamination,
the $K$ LFs of M giants, computed from the $(K, H-K)$ CMDs and assuming that M giants 
have $H-K < 0.6$, were computed, and the results are compared in Figure 8. 
The LFs were corrected statistically for foreground 
star contamination by combining the number counts of objects with 
$K < 15$ in the NGC 147 and NGC 185 fields, where the number of 
foreground stars is greatest, and then fitting a power-law to the 
counts as a function of $K-$magnitude. The power-law fit was then used to compute the 
number of foreground stars in each LF magnitude interval. 
A corresponding relation for NGC 205 was obtained by scaling the relation 
derived for NGC 147 and NGC 185 to match the number of bright foreground stars in NGC 205. 

	The LFs in Figure 8 have been normalized according to the number of stars with $K$ 
between 16.9 and 17.5 to facilitate galaxy-to-galaxy the comparisons. 
The LFs of M giants in NGC 185 and NGC 205 climb smoothly towards fainter 
$K-$magnitudes, with no obvious discontinuities. While the LFs of the outer regions of all 
three galaxies are in rough agreement, the inner regions of NGC 185 and NGC 205 clearly 
contain an excess number of stars with $K < 17$ when compared with NGC 147, and this is 
due to differences in the peak AGB brightness. To quantify these differences, 
the peak brightness at which the number counts, corrected for foreground star 
contamination, differ from zero at the $1-\sigma$ or higher level were 
computed, and the results are listed in the third column of Table 4. 
Important caveats are that the most evolved AGB stars (1) are undergoing rapid evolution, 
and so are relatively short-lived, and (2) can be photometric variables. Rapid 
evolution may cause the measured peak brightness to be underestimated, while there is 
a bias to identifying stars that are near the peak of their light curves as the brightest 
M giants, and this will act to overestimate the peak brightness. In addition, 
the peak brightnesses listed in Table 4 are for M giants, and in most regions 
there are brighter C stars; consequently, the measurements in Table 
4 are not AGB-tip brightnesses.

\section{COMPARING THE PROPERTIES OF M GIANTS AND C STARS}

	The $(M_K, H-K)$ and $(M_K, J-K)$ CMDs of the three galaxies, which form the 
basis for the subsequent analysis, are compared in Figures 9 and 10. The properties of M 
giants and C stars deduced from these data are discussed in the following sub-sections. 
The comparisons not only consider the overall appearance of the CMDs and LFs, but 
also examine the relative densities of M giants and C stars as a function of integrated 
near-infrared brightness.

\subsection{M Giants}

	The M giant sequences on the CMDs show clear galaxy-to-galaxy differences. These 
differences are perhaps most apparent among the peak M$_K$ brightnesses, which are also 
listed in the last column of Table 4. The peak of the M giant sequence is brightest in the 
inner regions of NGC 205, and is faintest in NGC 147 and the outer regions of NGC 185. 

	The ages of the M giants can be estimated by making comparisons with isochrones. 
In Figures 9 and 10 the CFHTIR observations are compared with 
Z = 0.008 log(t$_{yr}$) = 8.1, 9, and 10 isochrones from Girardi et al. (2002). 
Selecting a metallicity to compare with the observations is problematic, as there are no 
formal abundance estimates for the youngest stars in these galaxies. Nevertheless, 
Bica, Alloin, \& Schmidt (1990) use population synthesis techniques to find that 
the majority of stars with ages $< 1$ Gyr in NGC 205 have [M/H] $\leq -0.5$, while Butler 
\& Martinez-Delgado (2005) find that the most metal-rich stars in this galaxy have 
[Fe/H] $> -0.7$. Consequently, the Z = 0.008 models are adopted here to compare with the 
observations. The isochrones plotted in Figures 9 and 10 more-or-less bracket the range of 
M giant $J-K$ colors in each galaxy.

	The M giants with the bluest colors are in the 
inner regions of NGC 205, and the isochrones suggest that these 
stars have an age log(t$_{yr}) \sim 8.5$. For comparison, the brightest M giants in 
the inner regions of NGC 185 have an age log(t$_{yr}) \sim 9$, 
while the youngest M giants in NGC 147, which appear not to be centrally 
concentrated, have an age log(t$_{yr}) \sim 9.5$. Based on the peak M giant 
brightness, the outer regions of NGC 147 and NGC 185 have similar ages.

	The range of ages in each galaxy can be compared by examing the color 
distributions of the M giants. The histogram distributions of $J-K$ colors for 
stars with M$_K$ between --7.6 and --7.2 in the inner and outer regions of each galaxy 
are shown in Figure 11. This brightness range was selected because (1) the artificial star 
experiments indicate that the sample is complete in each galaxy in this brightness interval, 
and (2) stars spanning the full range of ages will be present; while the photometric 
errors will be smaller for intrinsically brighter stars, only a more restricted 
range of ages, that excludes the oldest populations, will be present. 

	The artificial star experiments indicate that the random 
errors in the NGC 205 data in this brightness interval are larger than those in the 
other two galaxies, and this complicates efforts to compare the color distributions. 
To correct for this, the color distributions of NGC 147 and NGC 185 
were convolved with a guassian to simulate their appearance as if they had the same 
random errors as in NGC 205. Another complication is that the galaxies have 
very different stellar densities. To compensate for this, the number of objects in each 
region have been scaled to match those expected from a system with M$_K = -16$ 
using the $K-$band surface brightness profile for each galaxy from the 2MASS Extended 
Source Catalogue (Jarrett et al. 2000). Figure 11 thus compares the `specific 
frequency' of M giants per unit $J-K$ interval. The specific 
frequencies of M giants and C stars selected according to other brightness and color 
criteria are the subject of other comparisons in this paper.

	The peak in the color distributions is due to M giants, and there is a red tail 
that extends past $(J-K)_0 = 1.5$ that is due to C stars. As is expected based on the 
comparison of the CMDs, there are marked differences between the color distributions of 
bright M giants in the inner regions of the three galaxies. NGC 185 and NGC 205 both contain 
a larger relative number density of stars with M$_K$ between --7.2 and --7.6 than in NGC 
147. NGC 205 contains a population of blue stars not seen in either of the other galaxies, 
as expected if it contains the youngest M giants in the sample. 
The colors of Z=0.008 isochrones from Girardi et al. (2002) at M$_K = -7.8$ 
are shown near the top of each panel, and these indicate that NGC 185 and NGC 
205 contain stars spanning a broad range of ages, with the youngest in 
NGC 205 having log(t$_{yr}) \sim 8$.

	The color distributions of stars in the outer regions of the galaxies are in much 
better agreement, although there is a blue population of giants in NGC 205 that 
is not seen in the outer regions of NGC 147 and NGC 185. The oldest M giants 
will have the reddest $(J-K)$ colors, and it is thus interesting that the specific 
frequencies of M giants in the outer regions of all three galaxies are in 
reasonable agreement when $(J-K)_0 \geq 1.3$. In fact, the comparison in Figure 11 
indicates that as one moves outside of the nuclear regions then there is 
improved galaxy-to-galaxy agreement among the relative densities of M giants with a 
given $J-K$, suggesting greater similarities in the stellar contents. It is also worth 
noting that the relative densities of C stars in the lower panel of Figure 11 are 
in good agreement, and this point is investigated at greater length in \S 5.2 using a larger 
sample of C stars.

	The bolometric luminosities of M giants provide another means of investigating 
the star-forming histories of these galaxies. $K-$band bolometric corrections, BC$_K$, 
were computed from the relation between BC$_K$ and $J-K$ for Galactic and LMC 
AGB stars given by Bessell \& Wood (1984), and these were then applied 
to the $K-$band brightnesses of stars with $J-K$ between 0.9 and 1.4. Bessell, 
Castelli, \& Plez (1998) use synthetic spectra generated from model atmospheres 
to investigate the color sensitivity of BC$_K$, and find that the 
empirical Bessell \& Wood (1984) relation agrees with the models to within $\pm 
0.1$ magnitude over a wide range of colors (e.g. Figure 20 of Bessell et al. 1998).

	The bolometric LFs of M giants in the galaxies are compared in Figure 12. 
Foreground stars near the bright end of the M giant sequence were rejected based on their 
location on the $(K, J-K)$ CMD. The LFs for the outer regions 
in this figure have been scaled to match the number of stars that would be expected 
if the outer region had the same integrated brightness as the inner region, based on the 
$K-$band light profiles from the 2MASS Extended Source Catalogue (Jarrett et al. 2000).

	The LFs of M giants in the inner and outer regions of NGC 147 are identical 
within their uncertainties, indicating that the M giant content of this 
galaxy is well mixed spatially, with no obvious centrally concentrated young 
population. The situation is very different in NGC 185 and NGC 205, where 
the LFs of M giants in the inner regions are flatter than in the outer regions, 
indicating that the inner regions of these galaxies contain a larger fraction of 
young AGB stars than the outer regions. There also may be a 
break in the LF of the inner region of NGC 205 between M$_{bol} = 
-5.5$ and --6, which could be a signature of a discontinuity in the star-forming history. 
Such breaks in the star-forming history of NGC 205 might be expected if 
star formation is triggered by interactions with M31 (Davidge 2003).

	The specific frequency of M giants is investigated 
further in Figure 13, where the bolometric LFs of the three galaxies have 
been scaled to match the counts expected in 
a system with M$_K = -16$, based on the $K-$band surface brightness profile for each 
galaxy from the 2MASS Extended Source Catalogue (Jarrett et al. 2000). There is 
general agreement among the specific frequencies of M giants near the faint end;  
this is perhaps not unexpected since as one goes fainter then stars produced over a broader 
range of ages are sampled, with the result that stochastic noise introduced by recent 
star forming events is suppressed and the number counts more closely track mass.

	Lee (1996) found that the bolometric LF of AGB stars in NGC 205 
is flatter than that of NGC 185, and this is confirmed in 
Figure 13. The comparisons in Figure 13 also reveal similarities in stellar 
content at the bright end, as the specific frequencies of M giants in the outer regions 
of NGC 147 and NGC 185 at all brightnesses probed by these data 
are in excellent agreement. This agreement is consistent with the 
similarity in AGB peak brightness and the mean $J-K$ color of M giants in 
the outer regions of these galaxies, discussed earlier.

\subsection{C Stars}

	The C star contents of the galaxies are compared in Figures 14 and 15, where 
the M$_K$ LFs of C stars, which are defined to be those objects with 
$(J-K)_0 < 1.4$ and M$_K < -7.25$ (\S 4), are shown.
Following the procedure used for Figure 12, the outer region LFs in Figure 14 have 
been scaled to match the integrated brightness in the corresponding inner region 
so that radial trends in C star content can be investigated. The comparisons in Figure 14 
suggest that the C stars in NGC 147 are uniformly distributed. On the other hand, 
the outer regions of NGC 185 and NGC 205 contain a higher number of C stars with M$_K \sim 
-7$ per unit $K-$band surface brightness than the inner regions. A similar trend can be seen 
in the mean M$_K$ values of C stars, which are shown in the second column of Table 5; 
the number of C stars upon which the means are based are listed in the third column 
of this table. The quoted uncertainties in the $<M_K>$ entries are 
$1-\sigma$ standard errors of the mean. The $<M_K>$'s in the inner and outer regions of 
NGC 147 differ at slightly more than the $1-\sigma$ level, whereas in NGC 185 and NGC 205 
the difference between $<M_K>$ in the inner and outer regions is significant at 
the $2-\sigma$ or higher level.

	The $<M_K>$ values in the outer regions of NGC 147 and NGC 185 are in reasonable 
agreement, and so the specific frequency of C stars in the outer regions of these 
galaxies might also be expected to agree. The specific frequency of C stars 
per 0.5 $K-$band magnitude interval is investigated in 
Figure 15, where the number counts have been scaled to 
match those expected in a stellar system with M$_K = -16$. 
Models predict that the brightest C stars will form in the youngest 
environments, and the comparisons in Figure 15 are consistent with this, 
as the number of C stars with M$_K = -8.5$ is higher 
in NGC 205 than in either NGC 185 and NGC 147. Whereas there are significant differences 
between the C star LFs in the upper panel of Figure 15, there is general agreement 
between the LFs of the outer regions in the lower panel of Figure 15, indicating 
that these galaxies have similar densities of C stars per unit $K-$band surface 
brightness outside of the nuclear regions.

	The bolometric LFs of C stars have also been investigated. 
$K-$band bolometric corrections, BC$_K$, were computed using the relation between 
BC$_K$ and $J-K$ for Galactic and LMC AGB stars from Bessell \& Wood (1984), and the 
resulting bolometric LFs are compared in Figures 16 and 17. As in Figures 12 and 14, the 
outer region LFs in Figure 16 have been scaled to match the integrated $K-$band brightness 
in the inner regions to allow possible radial trends in each galaxy to be investigated. 
The specific frequency of C stars per bolometric magnitude is investigated 
in Figure 17 by normalizing the counts to those expected from a source with M$_K = -16$.

	There is excellent agreement between the inner and outer region LFs of NGC 147 
in Figure 16, while there is clearly a difference between the LFs of the inner and outer 
regions of NGC 205. There is an obvious excess of C stars with M$_{bol} = -5$ in 
the inner regions of NGC 205 with respect to the other galaxies, as expected if 
the inner region of NGC 205 contains a larger fraction of the youngest 
C stars than in the other galaxies. However, it is apparent from the lower panel 
of Figure 17 that there is good agreement between the outer region LFs. 
The comparison in the lower panel of Figure 17 indicates that the specific 
frequency of C stars with a given M$_{bol}$ in the outer regions of these galaxies 
is remarkably similar, suggesting that these galaxies have evidently converted similar 
quantities of gas per unit integrated galaxy mass into stars during intermediate epochs.

	The C star contents of these galaxies have also been compared using 
two statistics that use the integrated brightnesses of C stars. The first statistic 
relies on the fact that whereas luminous M giants are an ubiquitous signature of AGB 
evolution over a wide range of ages, the peak production of C stars is 
delayed until systems have ages near $\sim 1$ Gyr (e.g. Maraston 1998). Consequently, 
the ratio of light from C stars to that from bright M giants is a measure of 
the relative star forming histories during intermediate ($\geq 1$ Gyr) and 
moderately recent ($\sim 0.1 - 1$ Gyr) epochs.

	The ratio of the total $K-$band light from C stars 
and M giants brighter than M$_K = -7.25$, $f^{C}_{AGB}$ was 
computed for these galaxies, and the results are listed in the middle column of Table 5. The 
uncertainties in this quantity due to statistical flucuations in the numbers of stars 
is roughly $\pm 0.03$. It is evident that $f^{C}_{AGB}$ varies within and between the 
galaxies, being lowest in the inner region of NGC 205, where the youngest 
stars in the three galaxies are located, and highest in NGC 147 and the 
outer region of NGC 185, where there is a lack of the very bright M giants seen in the 
nuclear regions of NGC 185 and NGC 205. Based on $f^{C}_{AGB}$, it appears that 
the relative star-forming histories of the outer regions of NGC 147 and NGC 185 
during intermediate and moderately recent epochs have been similar, in agreement with 
what would be inferred from previous comparisons.

	The most recent episodes of star formation in galaxies like NGC 185 and NGC 205 
likely involve only a modest fraction of the total galaxy mass, and so the integrated 
$K-$band light is expected to be dominated by much older populations. Consequently, the 
ratio of the total light from C stars to the total integrated light from a galaxy will 
provide insight into the star forming histories at intermediate and old 
epochs. To investigate such a dependence, the fraction of the total $K-$band 
light coming from C stars, $f^{C}_{Total}$, was computed using the surface brightness 
profile for each galaxy in the 2MASS Extended Source Catalogue, and the results are listed 
in the last column of Table 5. The estimated uncertainty in these entries, based on 
statistical flucuations in the number of C stars, is roughly $\pm 0.003$. 

	It is evident from Table 5 that $f^{C}_{Total}$ is roughly constant 
throughout these galaxies, with C stars contributing 2\% of the total $K-$band 
light; the standard deviation about the mean value 
of $f^{C}_{Total}$ is $\pm 0.004$, in rough agreement with what is expected from the 
uncertainties in $f^{C}_{Total}$. The system-to-system 
agreement in the last column of Table 5 argues that the 
star-forming histories of NGC 147, NGC 185, and NGC 205 
have likely been very similar when averaged over the time scales corresponding to 
peak C star production - i.e. over Gyrs.

\section{STELLAR VARIABILITY IN NGC 185 \& NGC 205}

	Based on studies of other systems, it can be anticipated that the majority 
of the brightest AGB stars in these galaxies will likely be LPVs. Indeed, the brightest 
AGB stars observed at any given time in a galaxy will almost certainly be 
LPVs near the peak of their light curves. NGC 185 and NGC 205 were both observed with the 
CFHTIR during an observing run in June 2001, and these data can be compared with the 
November 2002 observations to gain insight into variability among the brightest 
resolved stars. 

	Davidge \& Rigaut (2004) used 
$K-$band observations of the brightest stars in M32 recorded on two 
epochs to investigate the nature of variable stars in that galaxy, and the procedures 
used in that study are employed here. The key parameter 
is the difference in $K$ magnitude between the two epochs, 
$\Delta K$. The histogram distribution of $\Delta K$ contains information about the 
number of variables and the amplitude of their photometric variations. 

	The $K-$band brightnesses of stars in the June 2001 CFHTIR data were 
measured with the same procedures that were applied to the November 2002 data, 
and $\Delta K$ was computed by taking the difference between the two sets of brightnesses.
The histogram distributions of $\Delta K$ for stars with M$_K$ between 
--7.5 and --8.0 in the inner regions of NGC 185 and NGC 205 are shown in 
Figure 18; the properties of variable stars in the outer regions of these galaxies were 
not investigated because of the small number of objects in the overlapping areas 
of the two datasets. The dotted lines in Figure 18 show the distributions that would be 
expected if there were no variable stars, based on the 
errors in the photometry determined from the artificial 
star experiments. The $\Delta K$ distribution in the inner 
regions of NGC 185 is markedly wider than the predicted photometric 
uncertainties, indicating that some fraction of the brightest 
AGB stars are variable. For comparison, the $\Delta K$ distribution for NGC 205 roughly 
matches the scatter due to random photometric errors. As demonstrated below, this does not 
mean that LPVs are abscent in NGC 205; however, the relatively large photometric errors 
obviously confound efforts to probe the LPV content of this galaxy.

	Hughes \& Wood (1990) investigated the light curves of luminous AGB variables 
in the LMC, and the photometric measurements presented in 
that study were used to construct a reference 
$\Delta K$ distribution for comparison with the NGC 185 and NGC 
205 data. Following the procedure described by Davidge \& Rigaut 
(2004), individual measurements from Hughes \& Wood (1990) for LPVs 
observed in successive observing seasons were paired and $\Delta K$ was 
then computed for each pair; the pairing of points 
between successive seasons crudely replicates the time baseline between the two CFHTIR 
observing runs. Hughes \& Wood (1990) group the LMC stars into 
large and small amplitude LPVs. However, the standard deviations in the $\Delta K$ values 
obtained from the entries in their Tables II and III are similar ($\pm 
0.42$ versus $\pm 0.49$ mag), and so the $\Delta K$ values for both groups of LPVs 
were combined to create a single reference distribution. 

	To simulate the effects of observational scatter, the reference $\Delta K$ 
distribution constructed from the LMC LPVs was convolved with a gaussian, the standard 
deviation of which matched the dispersion measured from the 
artificial star experiments. In addition, some fraction of 
bright AGB stars are not variable. To account for these objects, a population 
of non-variable sources, amounting to a fraction $f_{nvar}$ of the total number of stars, 
was modelled as a gaussian distribution with a standard deviation 
matching the scatter predicted from the artificial star experiments. Non-variable 
components spanning a range of $f_{nvar}$ values were added to the raw reference 
distributions to create a suite of reference $\Delta K$ distributions.

	A best-fitting value of $f_{nvar}$ was found by comparing the observed and modelled 
$\Delta K$ distributions and finding the $f_{nvar}$ that minimizes the 
Kolomogorov-Smirnov D$_{max}$ statistic. The best-fitting models are compared with 
the observed distributions in Figure 18. The value of $f_{nvar}$ that gave the best 
match with the observations is listed in the upper left hand corner of each panel.

	When applied to the NGC 185 data, the fitting procedure described above indicates 
that $f_{nvar} = 0.3 \pm 0.15$ ($1-\sigma$ uncertainty). 
The agreement between the observed and modelled $\Delta K$ distribution 
in the upper panel of Figure 18 is not ideal. The poor agreement between the 
observed and modelled distributions when $\Delta K < -0.5$ suggests that the LPVs 
in NGC 185 may have a smaller amplitude distribution than those in the LMC, although 
there is better agreement between the models and observations at positive values of 
$\Delta K$. Models with $f_{nvar} < 0.3$ will better 
match the NGC 185 distribution near $\Delta K = 0$, although at the expense of the 
fit at very negative values of $\Delta K$. These problems notwithstanding, the value 
of $f_{nvar}$ computed for the inner regions of NGC 185 is consistent with what 
is seen in other galaxies, such as M32 (Davidge 
\& Rigaut 2004) and NGC 5128 (Rejkuba et al. 2003).

	The value of $f_{nvar}$ in NGC 205 is poorly constrained because of the 
large scatter in the photometric observations; the relatively large errors with respect to 
those seen in NGC 185 are a consequence of the greater distance and higher degree of 
crowding near the center of NGC 205. While the best fitting model has $f_{nvar} = 1.0$, 
other values of $f_{nvar}$ yield D$_{max}$ values that are not greatly different. Indeed, 
$f_{nvar} = 0$ can not be ruled out at the $2-\sigma$ significance level. 

\section{DISCUSSION \& SUMMARY}

	Moderately deep $J, H,$ and $K'$ images have been used to study the brightest AGB 
stars near the centers of the Local Group dE galaxies NGC 147, NGC 185, and NGC 205. The 
data sample stars with M$_K < -6$, and so are restricted to objects that are brighter than 
the RGB-tip. The main purpose of this investigation has been to compare 
the observational properties of the brightest oxygen-rich M giants and C 
stars in these galaxies, and the results are discussed below.

\subsection{Comparisons with Previous Studies}

	The AGB contents of NGC 185 and NGC 205 have been investigated previously 
in the near-infrared. Davidge (2003) used CFHTIR data to investigate 
the bright AGB content near the center of NGC 205 and 
found 320 C stars, whereas in the present study 387 C stars have been 
identified. The difference in the number of C stars is because Davidge (2003) adopted 
a redder $J-K$ color to differentiate between M giants and C stars; consequently, fewer 
C stars were found. The difference in this color criterion also 
affects the C star LFs, as AGB stars with bluer $J-K$ colors will tend to have lower 
luminosities. As a result, the C star LFs of NGC 205 computed here tend to be flatter at 
the faint end than those constructed by Davidge (2003).

	Kang et al. (2005) applied the techniques discussed by Davidge (2003) to identify C 
stars in near-infrared images of NGC 185. A total of 73 C stars with a mean brightness 
$<M_K> = -7.9$ were identified. For comparison, 103 C stars have been identified in this 
galaxy in the present study. The difference in C star numbers is due to the 
color criterion used to identify C stars; Kang et al. (2005) adopt a redder $J-K$ 
color for distinguishing between M giants and C stars, and hence find fewer C stars. 

	The mean C star brightness measured in the current study is 
fainter than that measured by Kang et al. (2005), as the 
mean brightness of all C stars in the November 2002 dataset is $<M_K> = -7.7$, 
whereas Kang et al. (2005) find that $<M_K> = -7.9$. 
However, Kang et al. (2005) used a different distance modulus than that adopted here. 
Had Kang et al. (2005) also adopted the McConnachie et al. 
(2004) distance for NGC 185, then they would have found $<M_K> = -7.7$, in good 
agreement with what is computed here.

	Kang et al. (2005) give the co-ordinates, brightnesses, and colors of 
C stars that were identified from their data, making it possible to conduct a 
star-by-star comparison of photometric properties. Using the November 2002 NGC 185 data, 
$K$ brightnesses and $J-K$ colors were obtained for 62 of the 73 C stars listed 
by Kang et al. (2005). The mean differences between the brightnesses and colors of the 
two sets of C star measurements are $\Delta K = -0.12 \pm 0.05$ magnitudes, 
and $\Delta(J-K) = 0.21 \pm 0.04$ magnitudes. The differences are in the sense 
Kang -- Davidge, and the quoted uncertainies are the standard errors in the means.

	There is considerable scatter in the $\Delta K$ measurements, and the standard 
deviation about the mean $\Delta K$ value is $\pm 0.36$ magnitudes. The majority of the 
C stars have $K < 17$, and a comparison with the entries in 
Table 3 indicate that the standard deviation in $\Delta K$ is at least three times that 
expected due to random photometric uncertainties alone. The large standard deviation in the 
$\Delta K$ measurements indicates that the majority of the C stars in NGC 185 are likely 
LPVs, as might be expected based on the comparisons in \S 6.

	Comparisons can also be made with studies of the brightest AGB stars in 
these galaxies that are based on observations at wavelengths shortward of $1\mu$m. Consider 
the peak AGB luminosities. Davidge (1994) used $V$ and $I$ data to find that the AGB-tip 
in NGC 147 occurs near M$_{bol}^{peak} = -5$, and this is in reasonable agreement with what 
is found from the CFHTIR data, where the most luminous star has M$_{bol} \sim 
-5.3$. Lee (1996) found that M$_{bol}^{peak} = -5.7$ in NGC 205 
from $V$ and $I$ images, and noted that this is 0.7 mag brighter than in NGC 185. 
For comparison, the CFHTIR data indicate that M$_{bol}^{peak} = -5.9$ in NGC 205, 
and M$_{bol}^{peak} = - 5.5$ in NGC 185. Lee (1996) found that the LF of oxygen-rich 
AGB stars in NGC 205 is flatter than that of oxygen-rich AGB stars in NGC 185, 
and this is in agreement with what is found from the CFHTIR data.

\subsection{Comparing the Relative Frequencies of C Stars}

	The majority of previous studies of C stars in these galaxies have used broad- and 
narrow-band images recorded at wavelengths shortward of 1$\mu$m. 
These studies tend to cover large areas, with the result that 
the C star contents are dominated by objects outside of the nuclear regions. 
While these studies show that the overall appearance of the $I-$band C star LFs of the 
three dEs are not greatly different, they also find a range in 
the ratio of C stars to M giants, C/M, that is suggestive of 
galaxy-to-galaxy differences in the relative density of C stars. Nowotny et al. (2003) 
find that C/M in NGC 185 is 58\% that in NGC 147, and point out that 
this is contrary to what would be expected given that NGC 185 may be more metal-poor 
than NGC 147. For comparison, Battinelli \& Demers (2004a,b) find that the C/M ratio in 
NGC 185 is 71\% that in NGC 147, and derive absolute C/M ratios that are higher than in 
the Nowotny et al. (2003) study; in the case of NGC 185 the C/M ratio computed by 
Battinelli \& Demers (2004b) is almost twice that measured by Nowotny 
et al. (2003). Perhaps most significantly, Demers et al. (2003) find that C/M in 
NGC 205 is 38\% of that in NGC 147. 

	From a purely observational perspective, the C/M ratio is sensitive to the depth of 
the photometric measurements (e.g. Figure 14 of Brewer, Richer, \& Crabtree 1995) and the 
method used to identify M giants (e.g. Battinelli \& Demers 2005). 
Battinelli \& Demers (2005) re-computed the C/M ratios of NGC 147, NGC 185, and 
NGC 205 in a homogeneous way using data from their earlier studies, and found that C/M$ = 
0.30 \pm 0.03$ in NGC 147, $0.24 \pm 0.03$ in NGC 185, and $0.20 \pm 0.01$ in NGC 205. 
While these new C/M ratios show better galaxy-to-galaxy agreement than those computed 
previously, they suggest the frequency of C stars, when normalized to the number of M 
giants, is not a constant in these galaxies.

	Should the C star contents of the three dEs be similar if they have had 
similar star-forming histories? Models of the advanced 
stages of AGB evolution predict that the third dredge-up, 
which can bring an excess of C into the outer envelope of a star, occurs 
over a broader range of masses as metallicity decreases (e.g. Marigo, Girardi, \& Bressan 
1999); as a result, the number of C stars is expected to increase in systems with lower 
metallicities. There is empirical evidence to support this expectation, as it has long been 
known that the C star content of galaxies, as measured by the C/M ratio, depends on 
metallicity (e.g. Cook et al. 1986, Pritchet et al. 1987, and Battinelli \& Demers 2005), 
in the sense that the relative number of C stars increases with decreasing 
metallicity. However, a complicating factor is that the impact of metallicity on the 
occurence of the third dredge-up is not the only factor that affects the C/M ratio, as the 
number of M giants also decreases as metallicity decreases.

	The RGB sequences in NGC 147, NGC 185, and NGC 205 suggest that 
the old stars in these galaxies have metallicities that differ by at most a few tenths 
of a dex (\S 1). Adopting the [Fe/H] versus C/M relation in Figure 3 of 
Battinelli \& Demers (2005), then a difference in 0.3 dex in [Fe/H] 
translates into a factor of 3 change in the C/M ratios. Given that at least part of this 
change in the C/M ratio is driven by a change in the number of M giants, then the C star 
densities in the three dE galaxies might be expected to differ by at most a factor of 3 
if they have experienced similar star-forming histories.

	The methods used here to compare C star contents rely on integrated quantities 
involving light from the entire galaxy and the AGB, and so are less sensitive to metallicity 
effects. That being said, the $(J-K)_0 > 1.4$ criterion that is employed here to identify C 
stars will likely select most of the objects that would be identified in broad and 
narrow-band photometric surveys at wavelengths shortward of $1\mu$m (Figure 7 of Demers et 
al. 2002). The statistic defined in this study to compare the relative C star densities 
that is most closely related to C/M is $f^{C}_{AGB}$, which is the 
fraction of $K-$band light from the bright portions of the AGB that comes from C stars. 
The $f^{C}_{AGB}$ entries in Table 5 suggest that this 
statistic tracks the C/M ratio measurements made by Battinelli \& Demers (2005) 
in the sense that $f^{C}_{AGB}$ is highest in NGC 147 and lowest in NGC 205.

	Other methods have been defined here to measure the relative density of C 
stars, and these suggest that the three dE companions of M31 have similar C star 
densities outside of their nuclear regions. The fraction of the integrated $K-$band light 
that originates from C stars, f$^{C}_{Total}$, has been computed, and there is only 
modest galaxy-to-galaxy dispersion in this quantity, which is comparable to that expected 
from measurement errors alone. The specific frequency of C stars, 
where published $K-$band surface brightness measurements are used to 
compute the relative numbers of C stars per brightness interval normalized to 
an integrated brightness M$_K = -16$, indicates that there is excellent galaxy-to-galaxy 
agreement in the C star contents of these galaxies outside of 
their nuclear regions. Indeed, the comparisons in the lower panel of 
Figure 17 indicate that the specific frequencies of C stars in the main bodies of 
NGC 147, NGC 185, and NGC 205 agree to well within a factor of 2.

\subsection{The Source of Star-Forming Material During Intermediate Epochs}

	The galaxy-to-galaxy agreement between the specific frequencies of C stars suggests 
that the star formation rates in the three galaxies, when normalized to the total galaxy 
masses, have been comparable over time scales of a few Gyr. 
This has implications for how star formation proceeds in these galaxies, and the source 
of the gas and dust from which stars form; that all three galaxies turned the same 
fractional amount of gas into stars during intermediate epochs suggests that (1) similar 
processes regulate star formation in these systems, and/or (2) the gas 
that fueled star formation originated in carefully regulated reservoirs. 

	The three dEs do not contain large reservoirs of star forming material at the 
present day (\S 1). However, modest amounts of material to fuel star formation can be 
supplied by low and intermediate mass stars during the advanced phases of their 
evolution, and this mechanism could provide a supply of star forming material that scales 
with galaxy mass. However, can mass loss from evolved low and intermediate mass stars 
provide enough gas to explain the mass of stars that evidently formed in the dEs 
during intermediate epochs? To answer this question we consider NGC 185 
as an example, and start by estimating the mass of intermediate age stars in that galaxy.

	A lower limit to the mass of intermediate age stars in NGC 185 is estimated in two 
ways. The first approach considers the mass of the young nucleus in NGC 185. The $K-$band 
surface brightness profile of NGC 185 (Jarrett et al. 2000) indicates that the central 
cusp has M$_K \sim -14.3$. Butler \& Martinez Delgado (2005) find that the cusp has an 
age $4 \times 10^8$ year. The Leitherer et al. (1999) models with a Salpeter 
mass function predict that a simple stellar 
system with this age and integrated brightness has a mass $\sim 3 \times 
10^6$M$_{\odot}$. The presence of faint C stars indicates that there has been 
at least one other episode of star formation in the galaxy 
during intermediate epochs. Assuming that a similar amount of gas was turned into 
stars then the total mass of stars formed over the past few Gyr is at least 
$\sim 6 \times 10^6$M$_{odot}$. Assuming a star forming efficiency of 10\%, then the 
mass of gas required is at least $\sim 6 \times 10^7$M$_{\odot}$.

	A second lower limit to the size of the intermediate age population in NGC 185 
can be estimated from the properties of the observed AGB stars. Stars evolving 
on the AGB make the maximum contribution to the integrated light 
from a moderately metal-poor simple stellar system when log(t) $\sim 9$ (Maraston 1998); 
hence, a lower limit to the mass of stars that formed during intermediate epochs can be 
obtained by assuming that the AGB stars have an age of 1 Gyr. With 4\% of 
the light coming from AGB stars brighter than $K = -7.25$ (Table 5), then the 
results shown in Figure 11 of Maraston (1998) predict that stars that 
formed during intermediate epochs in NGC 185 account for at least 2.5\% of the total galaxy 
mass, assuming that the intermediate-age population has a mass-to-light ratio in the 
$K-$band that is 0.5 that of older populations (e.g. Leitherer et al. 1999). 
With an integrated brightness M$_K = -17.4$ (Jarrett 
et al. 2000) and an assumed $K-$band M/L ratio of $\sim 1$, then NGC 185 has a total 
mass of $\sim 2 \times 10^8 M_{\odot}$. The mass of stars that formed during intermediate 
epochs is then at least $\sim 5 \times 10^6 M_{\odot}$. Assuming a 10\% star-forming 
efficiency, then the mass of gas required to form these stars is at 
least $\sim 5 \times 10^7 M_{\odot}$, in fortuitous agreement with the lower limit 
computed in the previous paragraph. 

	The amount of mass that is injected into the ISM by evolved 
stars is very uncertain, and two estimates are considered here. 
Gallagher \& Hunter (1981) conclude that star formation in NGC 185 is fueled 
by mass from evolved stars, and derive a mass return rate of $1 - 2 \times 
10^{-3} M_{\odot}$ year$^{-1}$ for NGC 185. For comparison, assuming that 
the majority of gas from evolved stars is ejected into the 
ISM by planetary nebulae, Welch et al. (1996) compute a mass return rate 
of $1.8 \times 10^{-4} M_{\odot}$ year$^{-1}$ for NGC 185. 
Adopting these two estimates as upper and lower limits, then the 
amount of gas that will be returned into the ISM in NGC 185 by low and intermediate 
mass stars over 10 Gyr is $2 - 20 \times 10^6 M_{\odot}$. The upper limit of this range 
is a factor of $\sim 3$ lower than the minimum amount of gas needed to explain the mass 
of intermediate age stars computed above. However, given that Gallagher \& Hunter (1981) 
estimate a factor of 2 uncertainty in their mass return rates then the hypothesis that 
the stars formed in these galaxies during intermediate epochs did so from gas 
returned to the ISM by evolved stars can not be rejected at present. Extremely deep imaging 
studies in the disk of the dEs that reach the main sequence turn-off will help establish 
the age of the intermediate age population and the relative mass of 
this component with respect to the main body of stars. This information will allow for a 
more rigorous test of the source of star-forming material.

\subsection{Where Did Intermediate Age Stars in NGC 147, NGC 185, and NGC 205 Form?}

	In \S 5 it was demonstrated that the specific frequency of M giants and C stars 
outside of the central 1 arcmin of the three dEs are in good agreement; aside from 
the obvious central concentrations of moderately young stars near the centers of 
NGC 185 and NGC 205, the AGB contents of these galaxies appear to be smoothly 
distributed out to radii of at least a few arcminutes. There have been multiple episodes 
of nuclear star formation in NGC 205, and the time between star forming episodes 
is consistent with the orbital period around M31 (Davidge 2003). Consequently, it is 
reasonable to anticipate that tidal interactions trigger the collapse of 
gas into the central regions of this galaxy, where stars then form. 

	If star formation occurs near the centers of the dEs then the nuclear star clusters 
must be disrupted on time scales of $\sim 1$ Gyr, in order for the progenitors 
of the C stars to be distributed to larger radii. 
However, in order to be disrupted on this time scale then processes other than 
normal dynamical evolution must be in play. This can be demonstrated using NGC 185 as an 
example. The majority of the youngest stars and star-forming material in NGC 185 are located 
in a region that is roughly $\sim 40$ parsecs across. The velocity dispersion in this 
part of the galaxy is $\sim 30$ km/sec (Held et al. 1992), and so the crossing time in the 
star-forming region is $\sim 10^6$ years. If the young cluster contains $\sim 10^6 - 
10^7$ stars then Equation 8-1 of Binney \& Tremaine (1987) indicates that the 
relaxation time is $\sim 10^10 - 10^11$ years. If clusters typically 
survive for $\sim 10$ relaxation times (e.g. Kim, Morris, \& Lee 1999), then the 
evaporation time for the central cluster is $\sim 10^{11} - 10^{12}$ years. 
Thus, stars that formed in the NGC 185 nuclear cluster 
will not be distributed into the main body of NGC 185 via simple dynamical evolution. 
If the majority of stars that formed in NGC 185 
during the past few Gyr did so in the nucleus, then some other process must be 
involked to hasten their distribution to larger radii.

	Interactions with M31 are one plausible mechanism for mixing the stellar contents 
of these galaxies. Indeed, there are indications that NGC 205, M32, and M31 have interacted 
at some point in the past (Cepa \& Beckman 1988; Ibata et al. 2001; Choi et al. 2002; 
McConnachie et al. 2004). However, van den Bergh (1998) argues that NGC 147 and NGC 185 
are a gravitationally bound pair, and that they thus likely do not have plunging orbits 
that bring them close to M31, as the binary system would then be disrupted. 
Battinelli \& Demers (2004b) point out that there is no evidence for a 
tidal bridge between these galaxies, and that the separation assumed by 
van den Bergh (1998) may be underestimated. Indeed, the adopted 
distances for these galaxies listed in Table 1 indicate that they are separated by 
more than 40 kpc, whereas the projected separation discussed by van den Bergh (1998) 
is 10 kpc. Therefore, the orbits of NGC 147 and NGC 185 may bring them close enough 
to M31 to spur star formation.

\acknowledgements{It is a pleasure to thank the anonymous referee for a prompt 
report that greatly improved the manuscript.}

\parindent=0.0cm

\clearpage

\begin{table*}
\begin{center}
\begin{tabular}{cccc}
\tableline\tableline
NGC & E(B--V)\tablenotemark{a} & $\mu_0$\tablenotemark{b} & R$_{M31}$ \\
 & & & (kpc) \\
\tableline
147 & 0.18 & 24.15 & 144.9 \\
185 & 0.18 & 23.95 & 188.5 \\
205 & 0.06 & 24.58 & 41.7 \\
\tableline
\end{tabular}
\end{center}
\tablenotetext{a}{From Schlegel et al. (1998)}
\tablenotetext{b}{From McConnachie et al. (2004)}
\caption{The Adopted Reddenings and Distance Moduli}
\end{table*}

\clearpage

\begin{table*}
\begin{center}
\begin{tabular}{clccc}
\tableline\tableline
NGC & & $\sigma_K$ & $\sigma_{HK}$ & $\sigma_{JK}$ \\
\tableline
147 & Inner & $\pm 0.12$ & $\pm 0.07$ & $\pm 0.09$ \\ 
 & Outer & $\pm 0.11$ & $\pm 0.07$ & $\pm 0.09$ \\
 & & & & \\
185 & Inner & $\pm 0.13$ & $\pm 0.06$ & $\pm 0.09$ \\ 
 & Outer & $\pm 0.11$ & $\pm 0.06$ & $\pm 0.07$ \\
 & & & & \\
205 & Inner & $\pm 0.15$ & $\pm 0.10$ & $\pm 0.11$ \\ 
 & Outer & $\pm 0.13$ & $\pm 0.09$ & $\pm 0.12$ \\
 & & & & \\
\tableline
\end{tabular}
\end{center}
\caption{Predicted Dispersion in the Photometry of an M Giant with K = 17}
\end{table*}

\clearpage

\begin{table*}
\begin{center}
\begin{tabular}{clccc}
\tableline\tableline
NGC & & $\sigma_K$ & $\sigma_{HK}$ & $\sigma_{JK}$ \\
\tableline
147 & Inner & $\pm 0.12$ & $\pm 0.20$ & $\pm 0.15$ \\ 
 & Outer & $\pm 0.11$ & $\pm 0.11$ & $\pm 0.13$ \\
 & & & & \\
185 & Inner & $\pm 0.13$ & $\pm 0.10$ & $\pm 0.09$ \\ 
 & Outer & $\pm 0.11$ & $\pm 0.10$ & $\pm 0.11$ \\
 & & & & \\
205 & Inner & $\pm 0.15$ & $\pm 0.20$ & $\pm 0.22$ \\ 
 & Outer & $\pm 0.13$ & $\pm 0.19$ & $\pm 0.17$ \\
 & & & & \\
\tableline
\end{tabular}
\end{center}
\caption{Predicted Dispersion in the Photometry of a C Star with K = 17}
\end{table*}

\clearpage

\begin{table*}
\begin{center}
\begin{tabular}{clcc}
\tableline\tableline
NGC & & $K_{Peak}$ & M$_{K}^{Peak}$\\
\tableline
147 & Inner & 16.6 & --7.6 \\
 & Outer & 16.4 & --7.8 \\
 & & & \\
185 & Inner & 15.8 & --8.2 \\
 & Outer & 16.4 & --7.6 \\
 & & & \\
205 & Inner & 15.6 & --9.0 \\
 & Outer & 16.2 & --8.4 \\
\tableline
\end{tabular}
\end{center}
\caption{Peak M Giant Brightness}
\end{table*}

\clearpage

\begin{table*}
\begin{center}
\begin{tabular}{ccccc}
\tableline\tableline
Galaxy & $<M_K>$\tablenotemark{a} & n$_C$\tablenotemark{b} & $f^{C}_{AGB}$\tablenotemark{c} & $f^{C}_{Total}$\tablenotemark{d} \\
\tableline
NGC 147  & & & & \\
Inner & $-7.63 \pm 0.04$ & 41 & 0.49 & 0.018 \\
Outer & $-7.71 \pm 0.05$ & 24 & 0.43 & 0.020 \\
 & & & & \\
NGC 185 & & & & \\
Inner & $-7.76 \pm 0.03$ & 74 & 0.30 & 0.022 \\
Outer & $-7.66 \pm 0.04$ & 29 & 0.43 & 0.018 \\
 & & & & \\
NGC 205 & & & & \\
Inner & $-7.97 \pm 0.03$ & 243 & 0.25 & 0.029 \\
Outer & $-7.86 \pm 0.03$ & 144 & 0.30 & 0.025 \\
\tableline
\end{tabular}
\end{center}
\tablenotetext{a}{Mean C star brightness.}
\tablenotetext{b}{Number of C stars.}
\tablenotetext{c}{Fractional $K-$band contribution made by C stars to the total AGB flux.
The uncertainty in this quantity based on statistical flucuations in the number of 
stars is roughly $\pm 0.03$.}
\tablenotetext{d}{Fractional $K-$band contribution made by C stars to the integrated galaxy 
light. The uncertainty in this quantity based on statistical flucuations in the number of 
C stars is roughly $\pm 0.003$.}
\caption{C star Statistics}
\end{table*}

\clearpage

\begin{figure}
\figurenum{1}
\epsscale{1.0}
\plotone{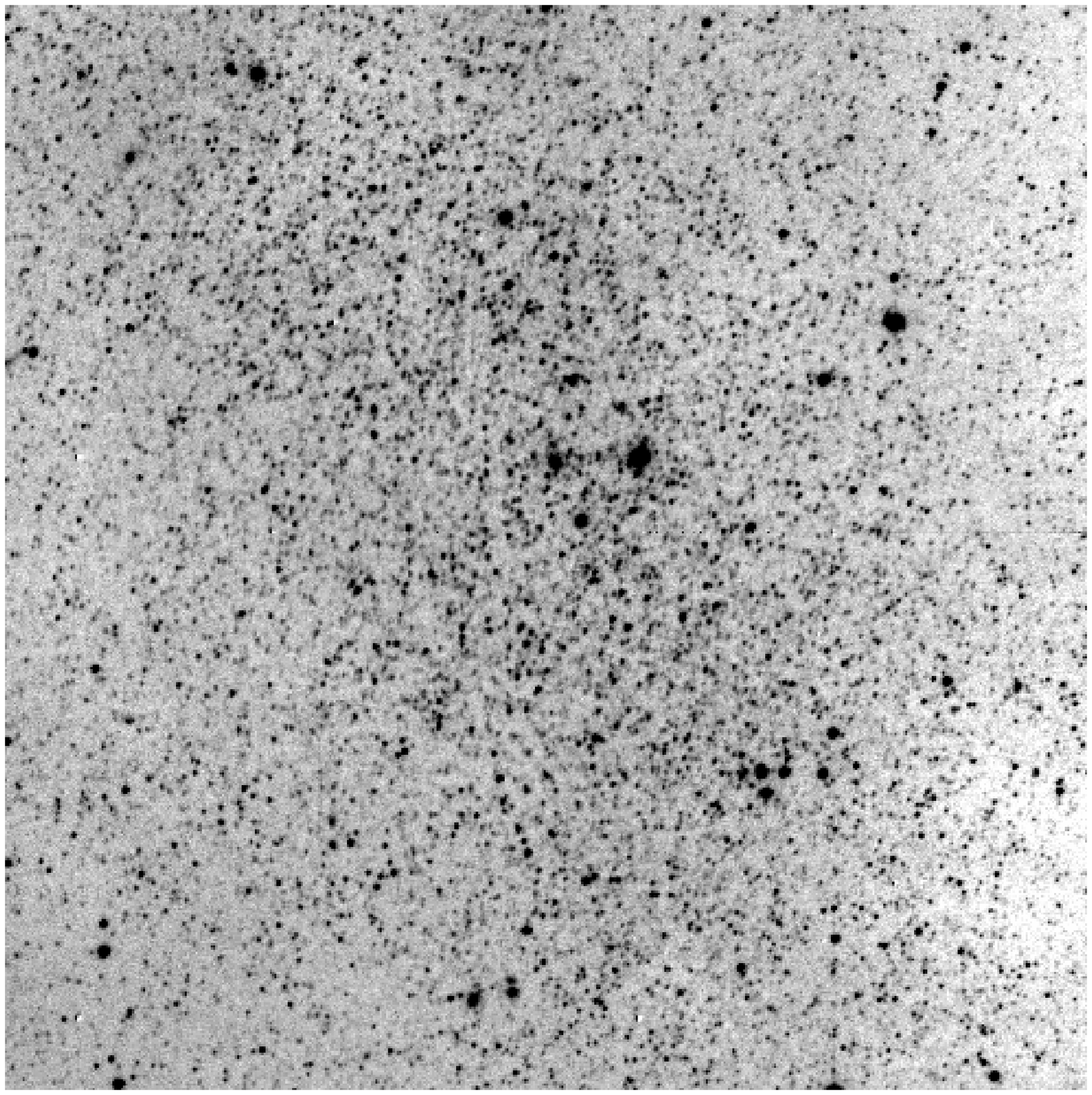}
\caption
{The central $3.7 \times 3.7$ arcmin$^2$ of NGC 147, as imaged in $J$
with the CFHTIR. North is at the top, and east is to the left.}
\end{figure}

\clearpage

\begin{figure}
\figurenum{2}
\epsscale{1.0}
\plotone{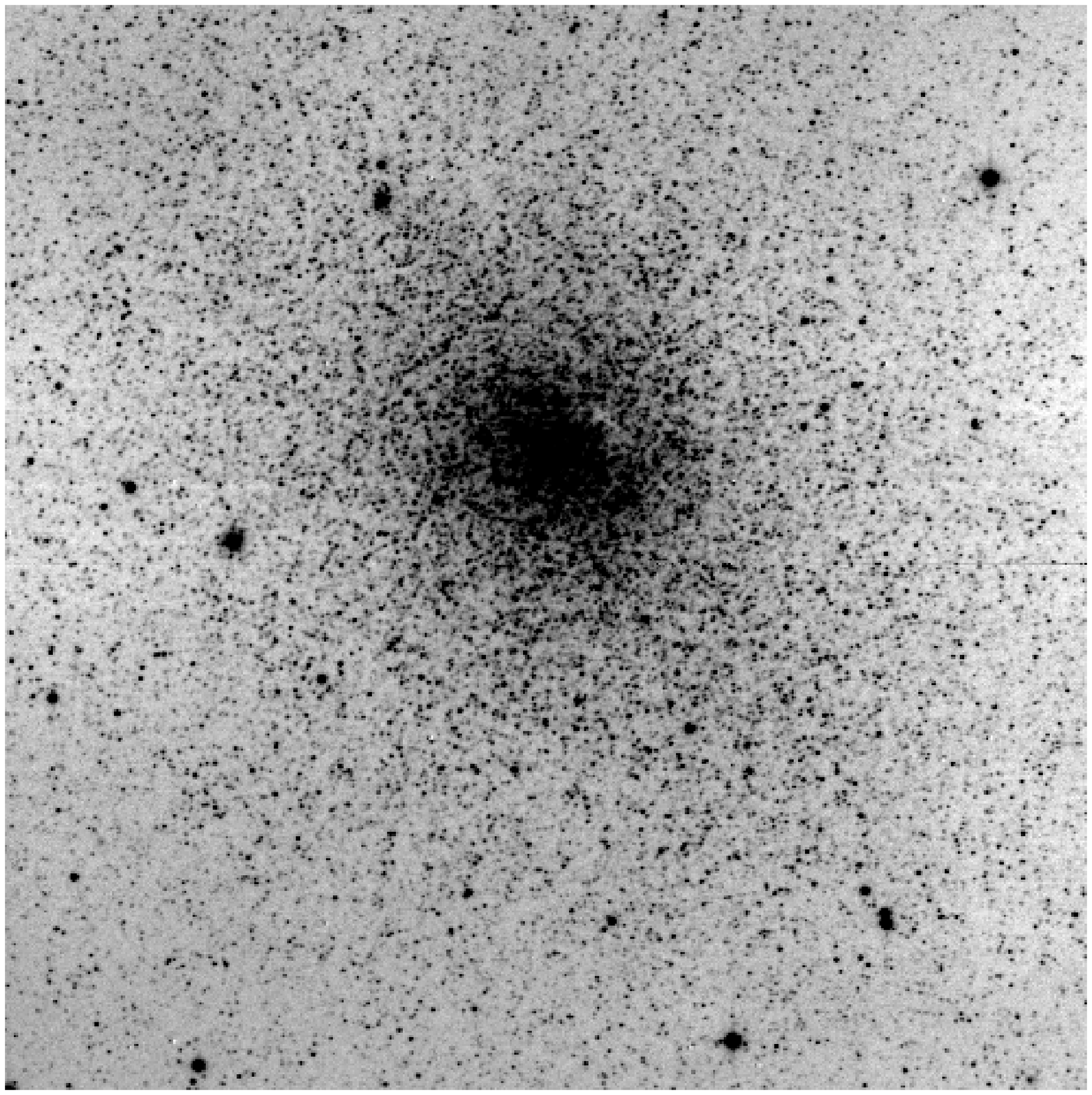}
\caption
{The central $3.7 \times 3.7$ arcmin$^2$ of NGC 185, as imaged in $J$ 
with the CFHTIR. North is at the top, and east is to the left.}
\end{figure}

\clearpage

\begin{figure}
\figurenum{3}
\epsscale{1.0}
\plotone{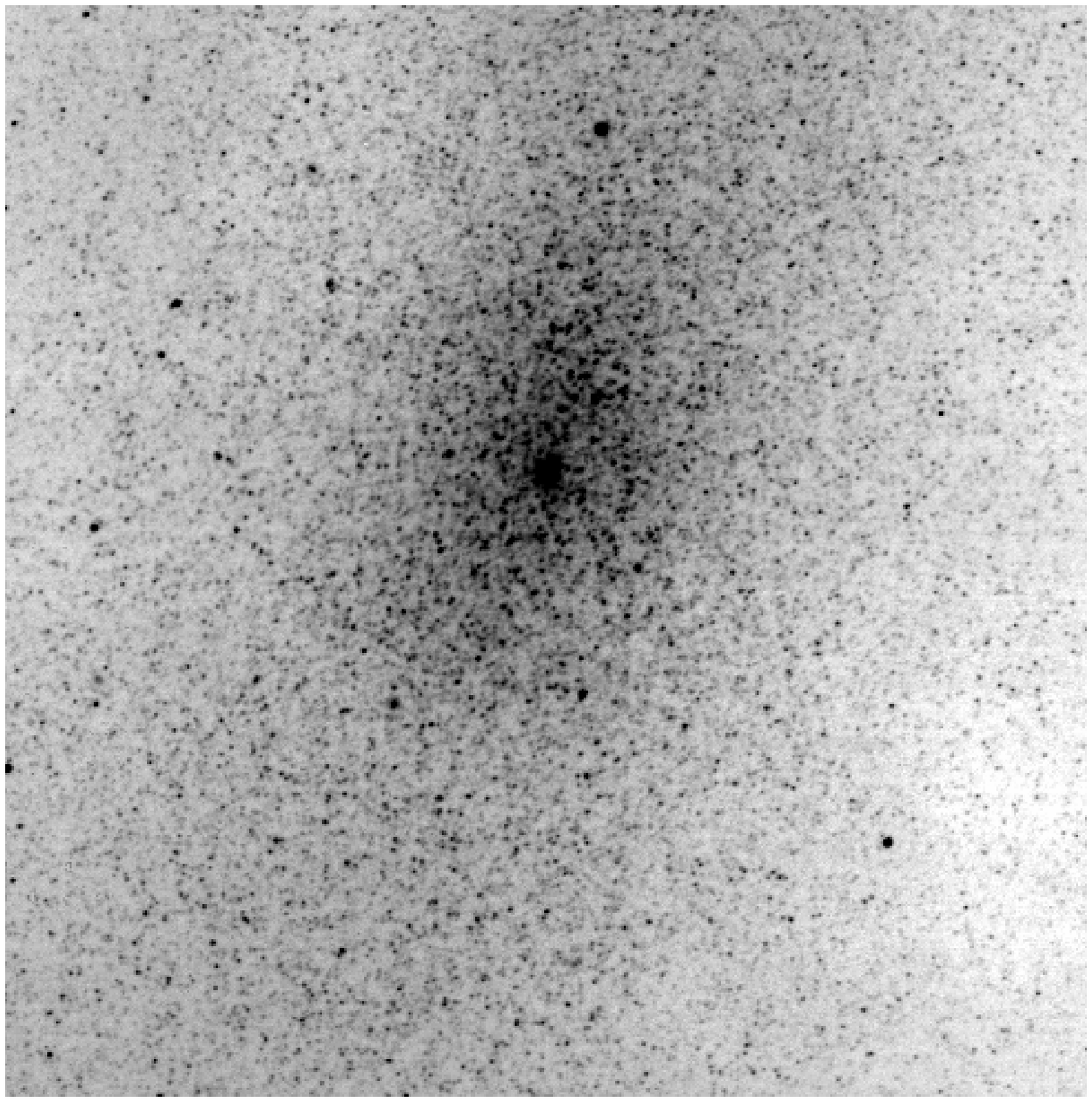}
\caption
{The central $3.7 \times 3.7$ arcmin$^2$ of NGC 205, as imaged in $J$ 
with the CFHTIR. North is at the top, and east is to the left.}
\end{figure}

\clearpage

\begin{figure}
\figurenum{4}
\epsscale{1.0}
\plotone{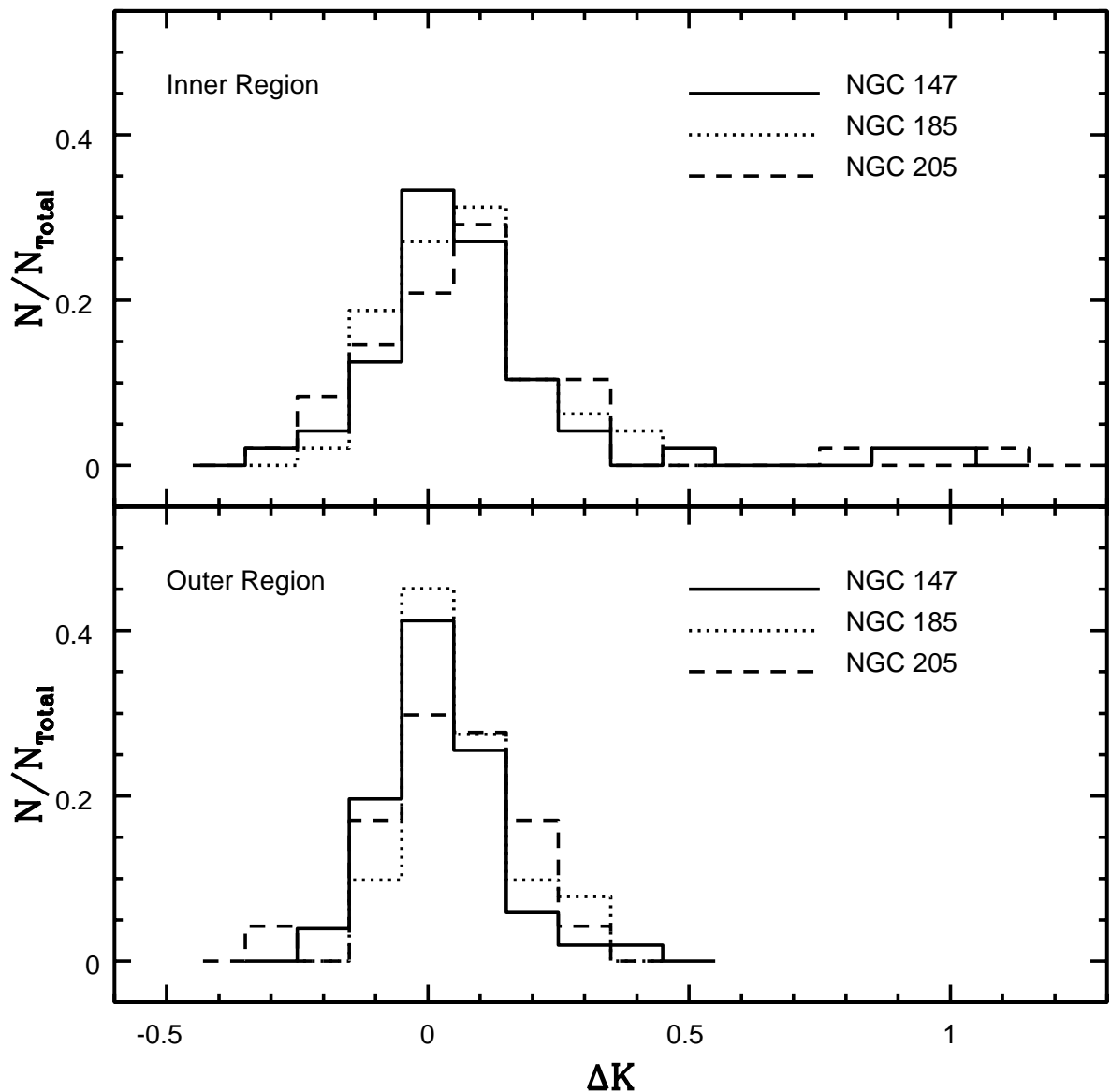}
\caption
{The histogram distribution of the difference between the actual and measured 
brightness of artificial stars, $\Delta$K. The artificial stars have $K = 17$ 
and colors consistent with M giants; N$_{0.1}$ is the number of stars per 
0.1 magnitude bin in $\Delta$K while N$_{Total}$ is the total 
number of artificial stars that have been recovered. Note that 
the $\Delta$K distributions are centered at or near $\Delta$K = 0, indicating that 
blending does not affect the photometry of the majority of objects with $K = 17$.}
\end{figure}

\clearpage

\begin{figure}
\figurenum{5}
\epsscale{1.0}
\plotone{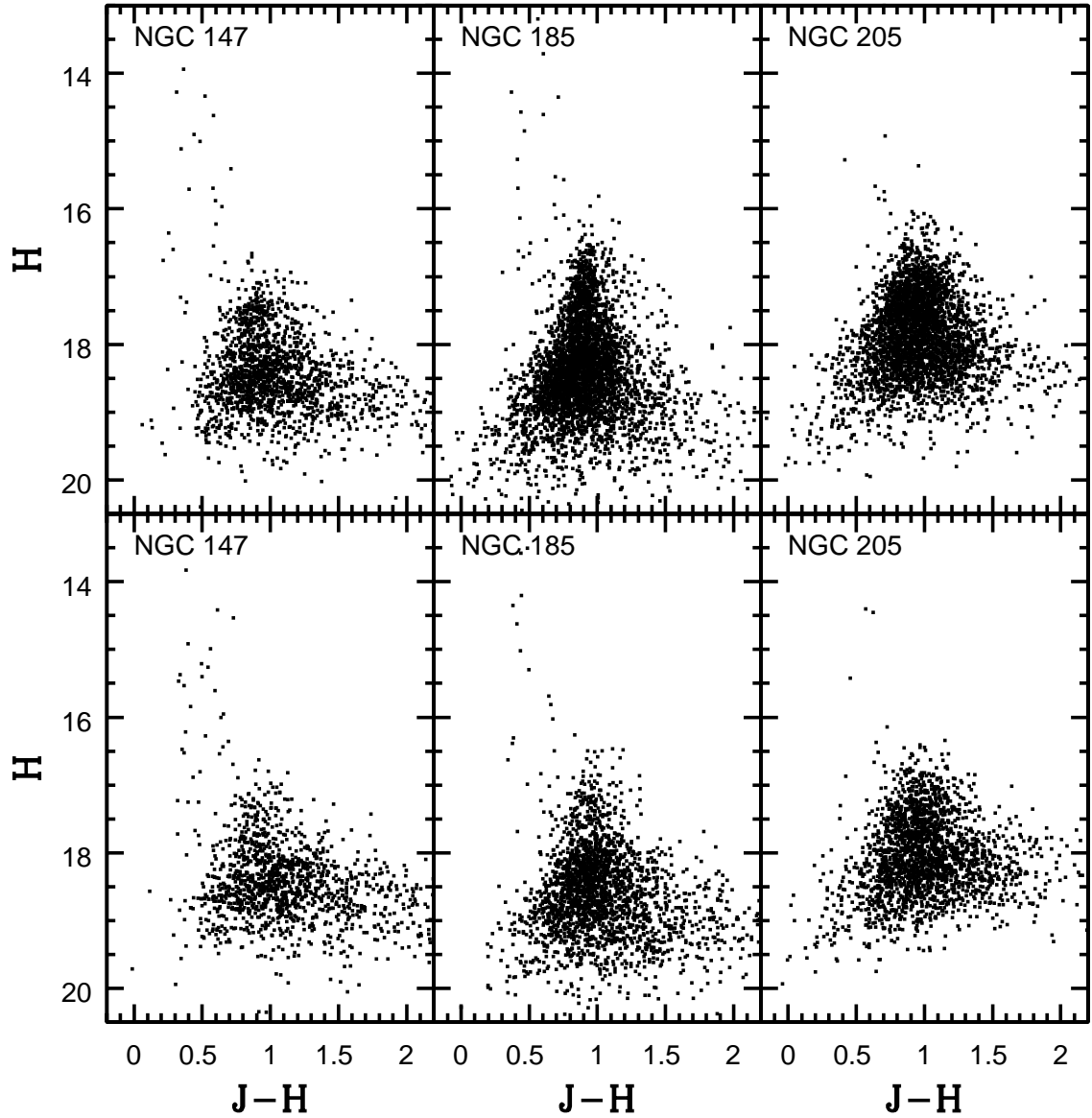}
\caption
{The $(H, J-H)$ CMDs of the inner and outer regions of NGC 147, NGC 185, and 
NGC 205. C stars form a sequence that plunges downward to the right 
of the dominant M giant sequence in each CMD.}
\end{figure}

\clearpage

\begin{figure}
\figurenum{6}
\epsscale{1.0}
\plotone{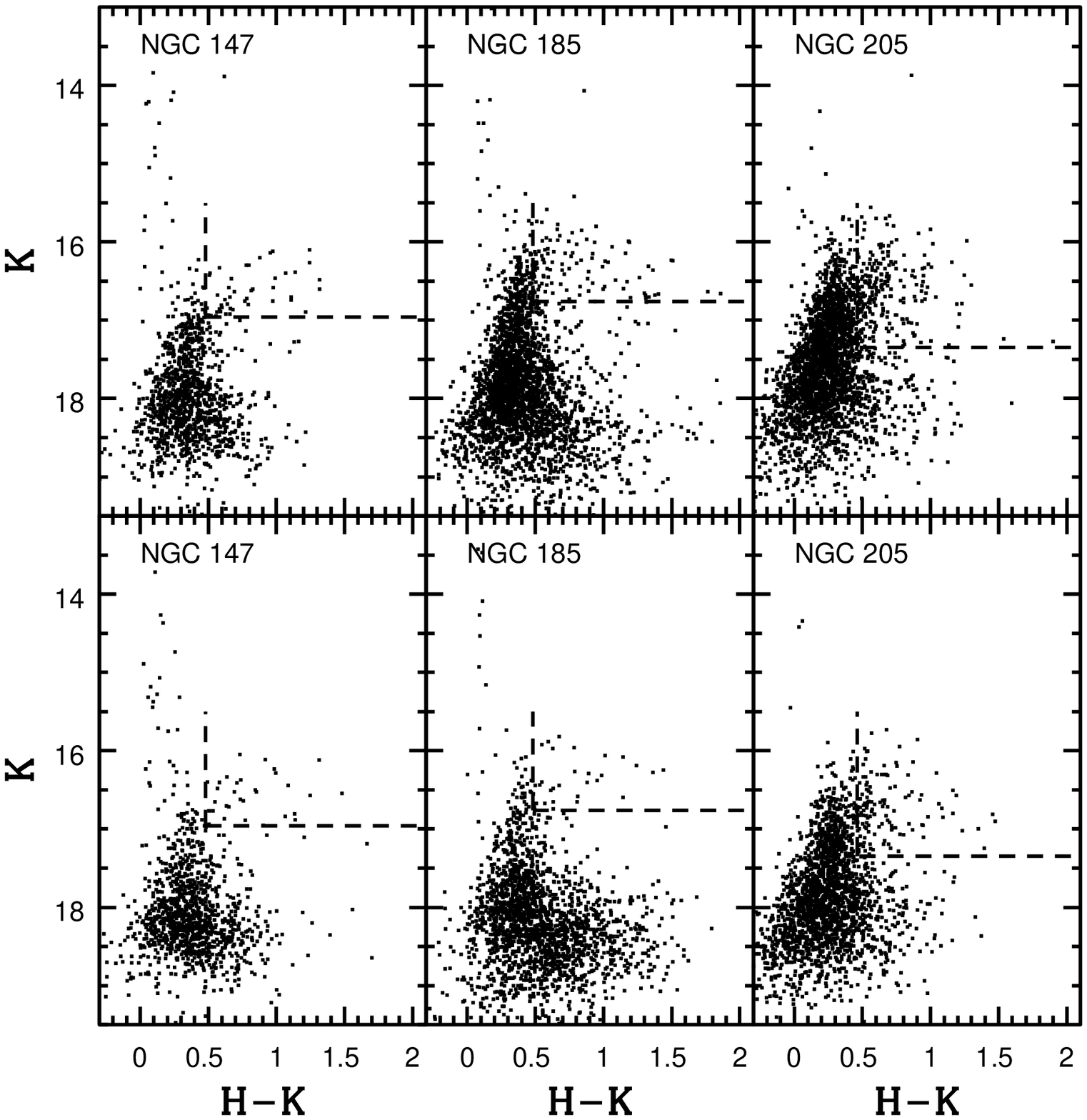}
\caption
{The $(K, H-K)$ CMDs of the inner and outer regions 
of NGC 147, NGC 185, and NGC 205. The dashed line marks the area containing 
C stars, which is bounded by $(H-K)_0 = 0.45$ and M$_K = -7.25$ (see text).} 
\end{figure}

\clearpage

\begin{figure}
\figurenum{7}
\epsscale{1.0}
\plotone{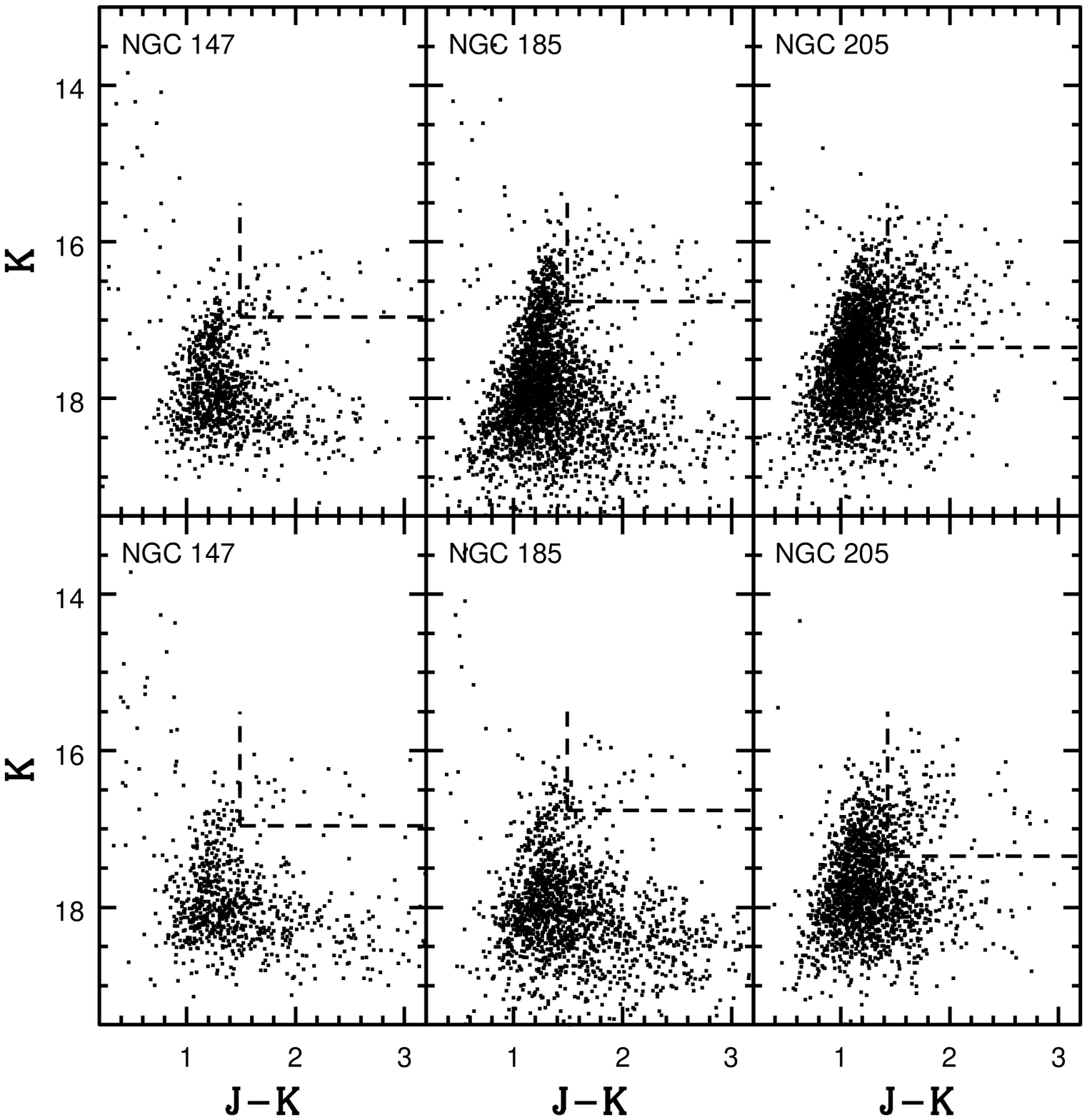}
\caption
{The $(K, J-K)$ CMDs of the inner and outer regions of 
NGC 147, NGC 185, and NGC 205. The dashed line marks the 
area containing C stars, which is bounded by $(J-K)_0 = 1.4$ and M$_K = -7.25$ (see text).} 
\end{figure}

\clearpage

\begin{figure}
\figurenum{8}
\epsscale{1.0}
\plotone{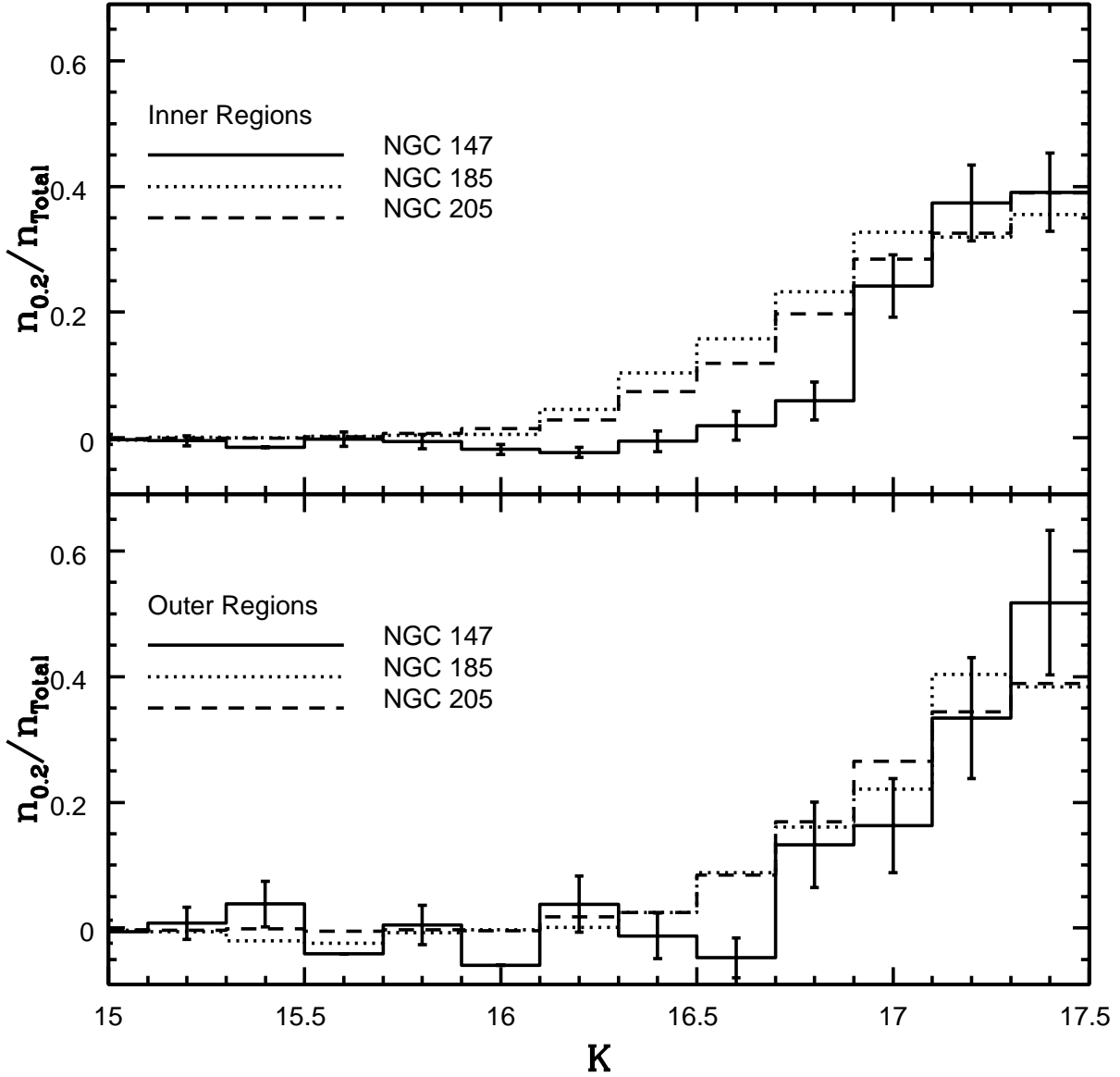}
\caption
{The $K$ LFs of M giants, which are assumed to have $(H-K)_0 < 0.45$. 
The LFs have been normalized according to the number of stars with $K$ between 16.9 and 
17.5; n$_{0.2}$ is the number of stars per 0.2 magnitude interval with $(H-K)_0 < 0.45$ in 
the $(K, H-K)$ CMD, while n$_{Total}$ is the total number of stars with $(H-K)_0 < 0.45$ 
and $K$ between 16.9 and 17.5. The errorbars show the $1-\sigma$ Poisson uncertainties 
in the NGC 147 measurements. The LFs have been corrected for 
foreground star contamination using the procedure described in the text.}
\end{figure}

\clearpage

\begin{figure}
\figurenum{9}
\epsscale{1.0}
\plotone{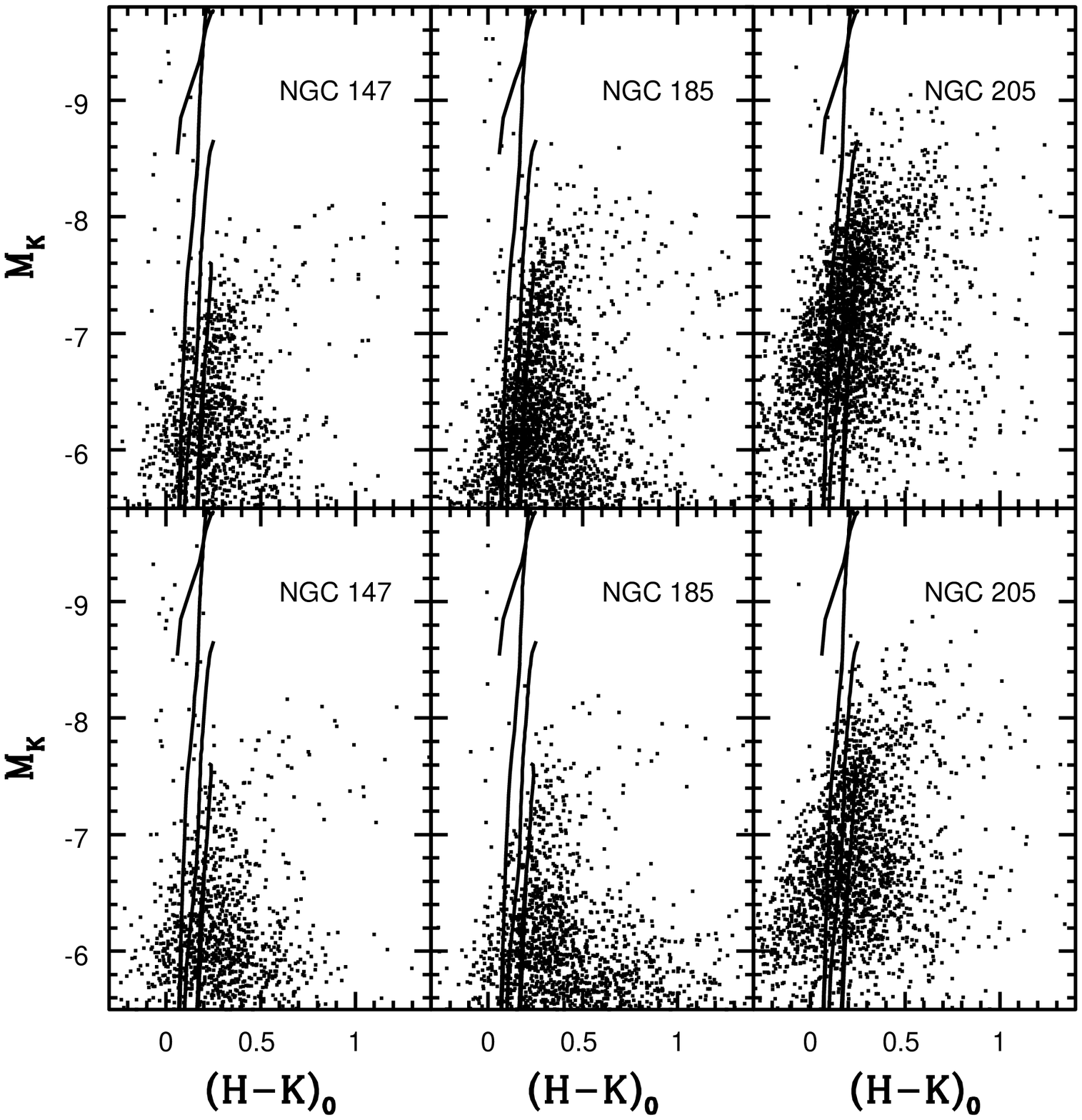}
\caption
{The $(M_K, H-K)$ CMDs of the inner and outer regions 
of NGC 147, NGC 185, and NGC 205. The solid lines are Z = 0.008 isochrones 
from Girardi et al. (2002) with log(t$_{yr}) =$ 8.1, 9.0, and 10.0.} 
\end{figure}

\clearpage

\begin{figure}
\figurenum{10}
\epsscale{1.0}
\plotone{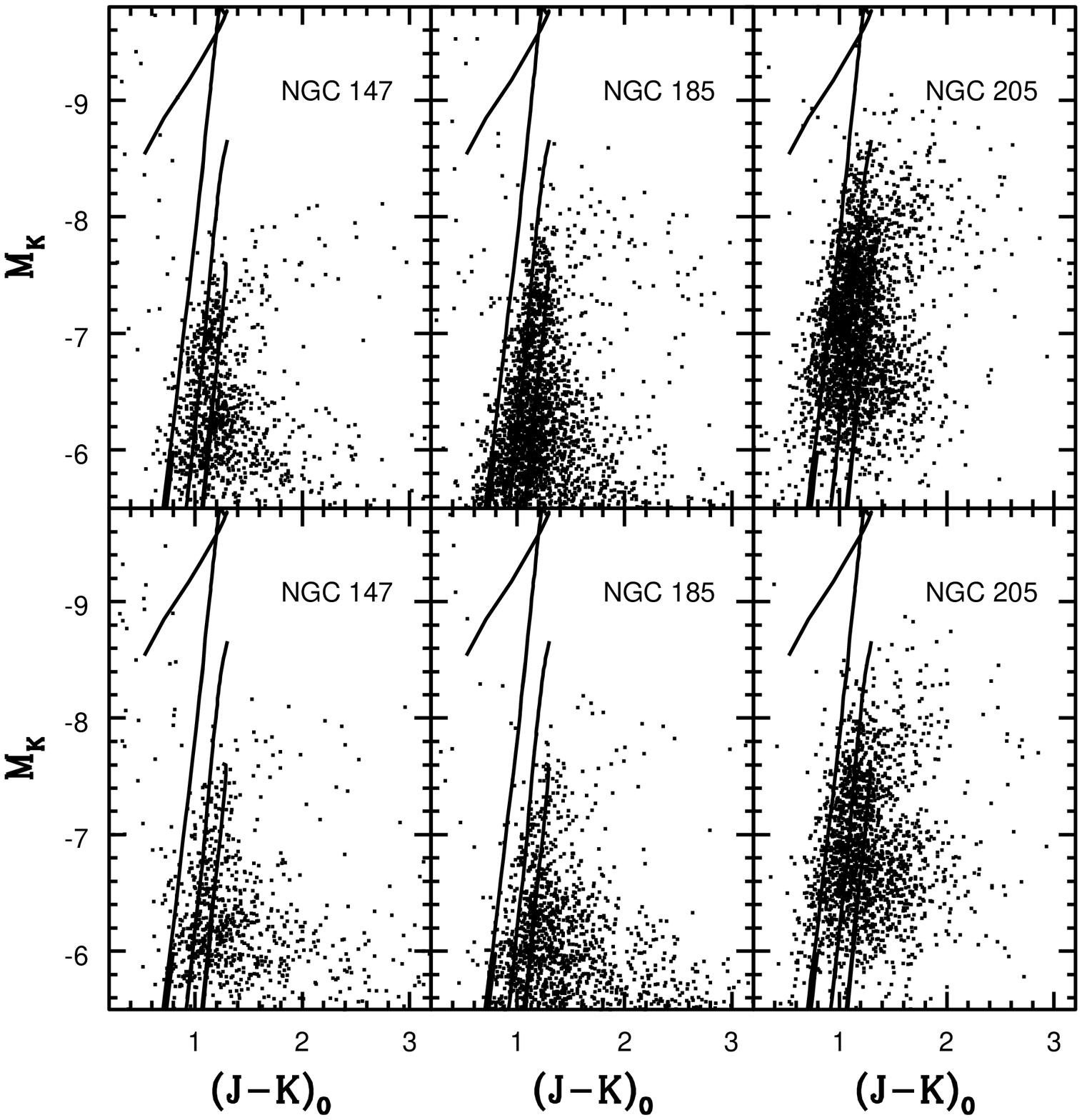}
\caption
{The $(M_K, J-K)$ CMDs of the inner and outer regions 
of NGC 147, NGC 185, and NGC 205. The solid lines are Z = 0.008 isochrones 
from Girardi et al. (2002) with log(t$_{yr}) =$ 8.1, 9.0, and 10.0.} 
\end{figure}

\clearpage

\begin{figure}
\figurenum{11}
\epsscale{1.0}
\plotone{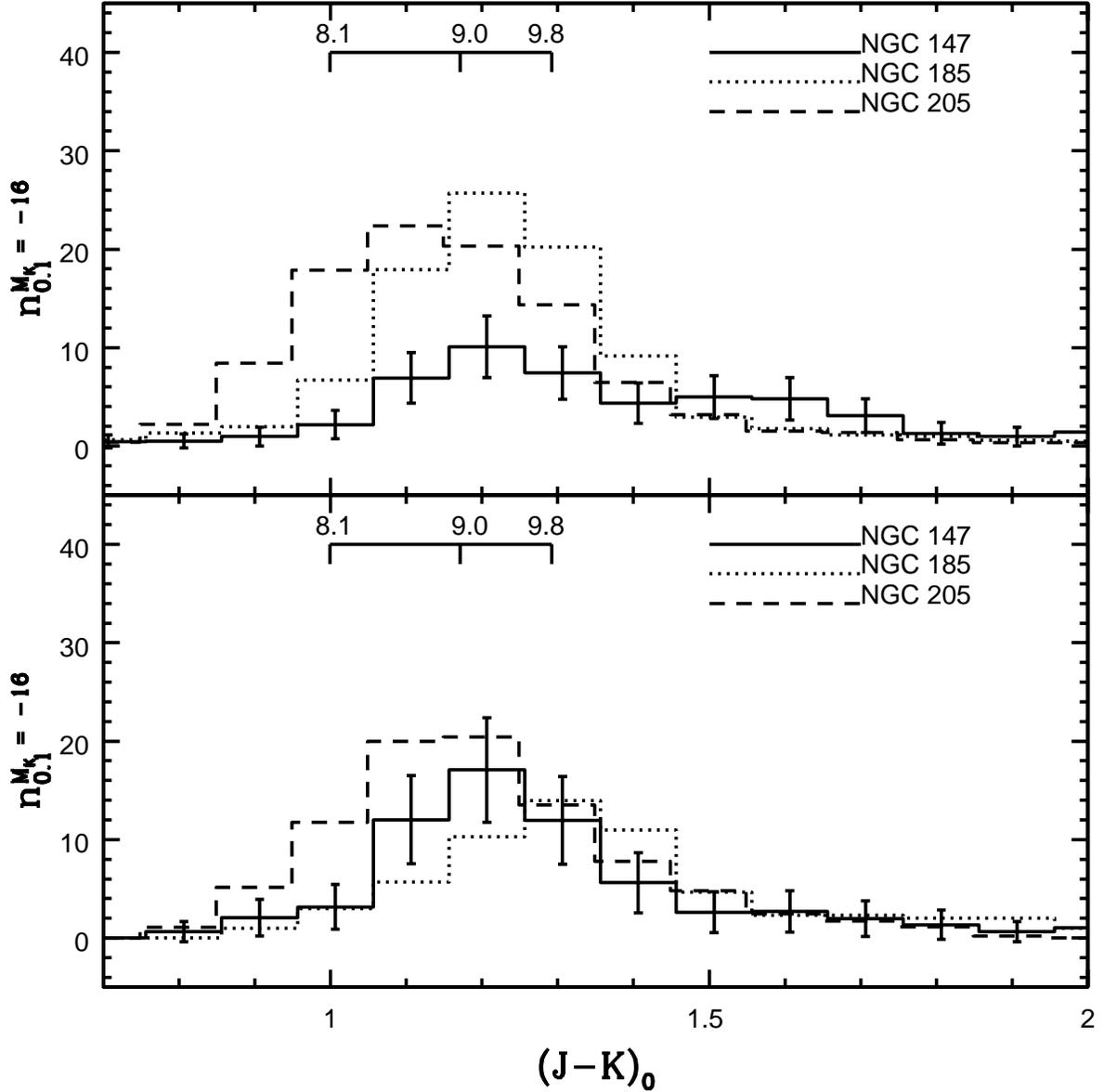}
\caption
{The histogram distribution of the $J-K$ colors of stars in the inner 
regions of NGC 147, NGC 185, and NGC 205; $n_{0.1}^{M_K = -16}$ is the number of stars 
with M$_K$ between --7.2 and --7.6 per 0.1 magnitude interval in $J-K$ for a population with 
M$_K = -16$. The distributions for NGC 147 and NGC 185 have been convolved with a 
gaussian to match the photometric uncertainties in the NGC 205 data. 
The errorbars in the number counts show $1-\sigma$ uncertainties in the 
NGC 147 data computed from Poisson statistics. The colors predicted by the Z=0.008 
Girardi et al. (2002) isochrones for M$_K = -7.4$ and 
ages log(t) = 8.1, 9.0, and 9.8 are also indicated.}
\end{figure}

\clearpage

\begin{figure}
\figurenum{12}
\epsscale{1.0}
\plotone{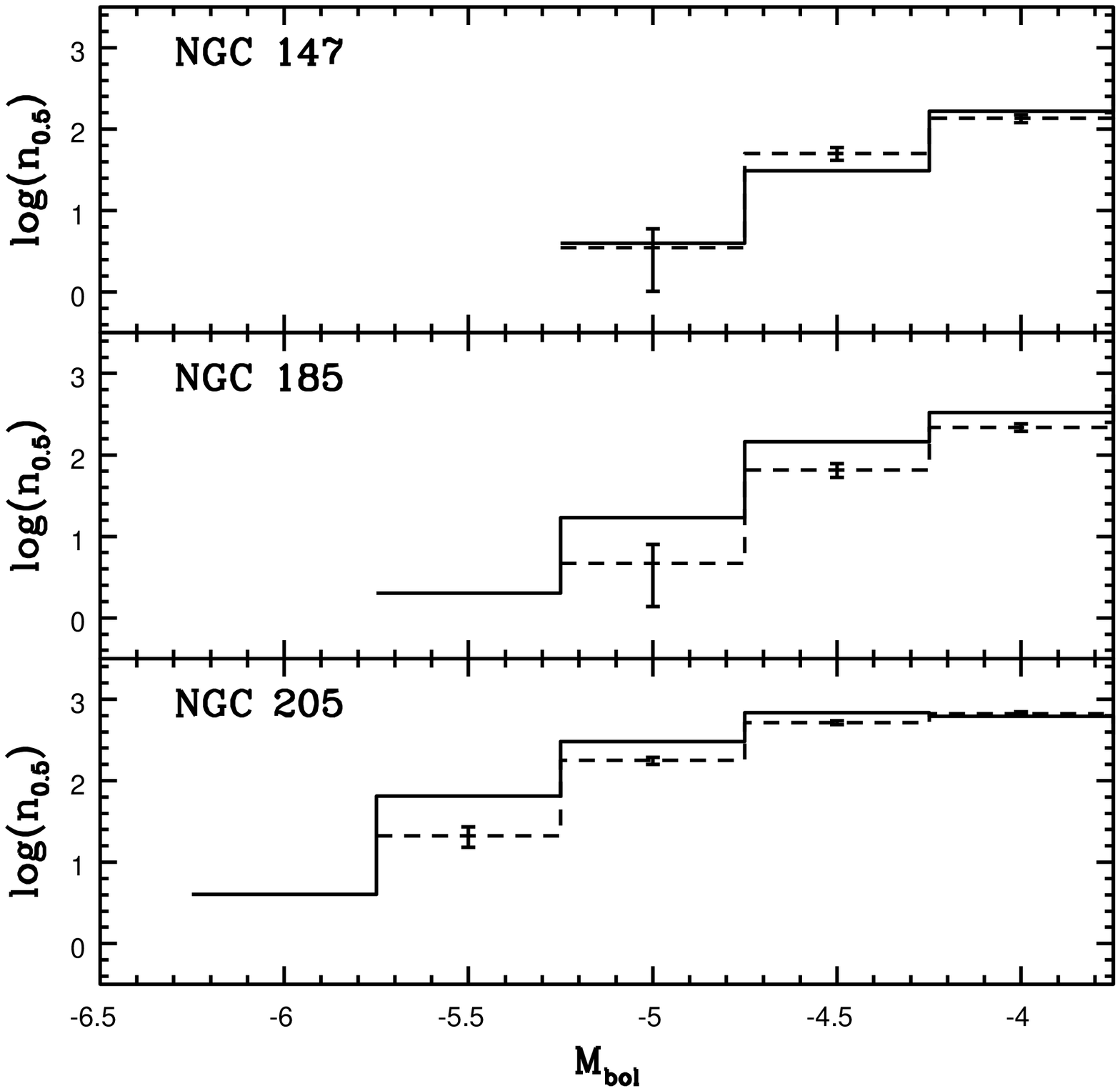}
\caption
{The bolometric LFs of M giants in the inner (solid lines) and outer (dashed lines) regions 
of NGC 147, NGC 185, and NGC 205; $n_{0.5}$ is the number of stars per 0.5 bolometric 
magnitude interval, and the errorbars show the $1-\sigma$ Poisson uncertainties 
in the outer region measurements. The number counts for the outer region data have been 
scaled to match the integrated $K-$band brightness in the inner region based on the $K-$band 
surface brightness profile in the 2MASS Extended Source Catalogue (Jarrett et al. 2000).
Bolometric corrections were computed using the calibration 
for oxygen-rich stars in the Milky-Way and the LMC from Bessell \& Wood (1984).}
\end{figure}

\clearpage

\begin{figure}
\figurenum{13}
\epsscale{1.0}
\plotone{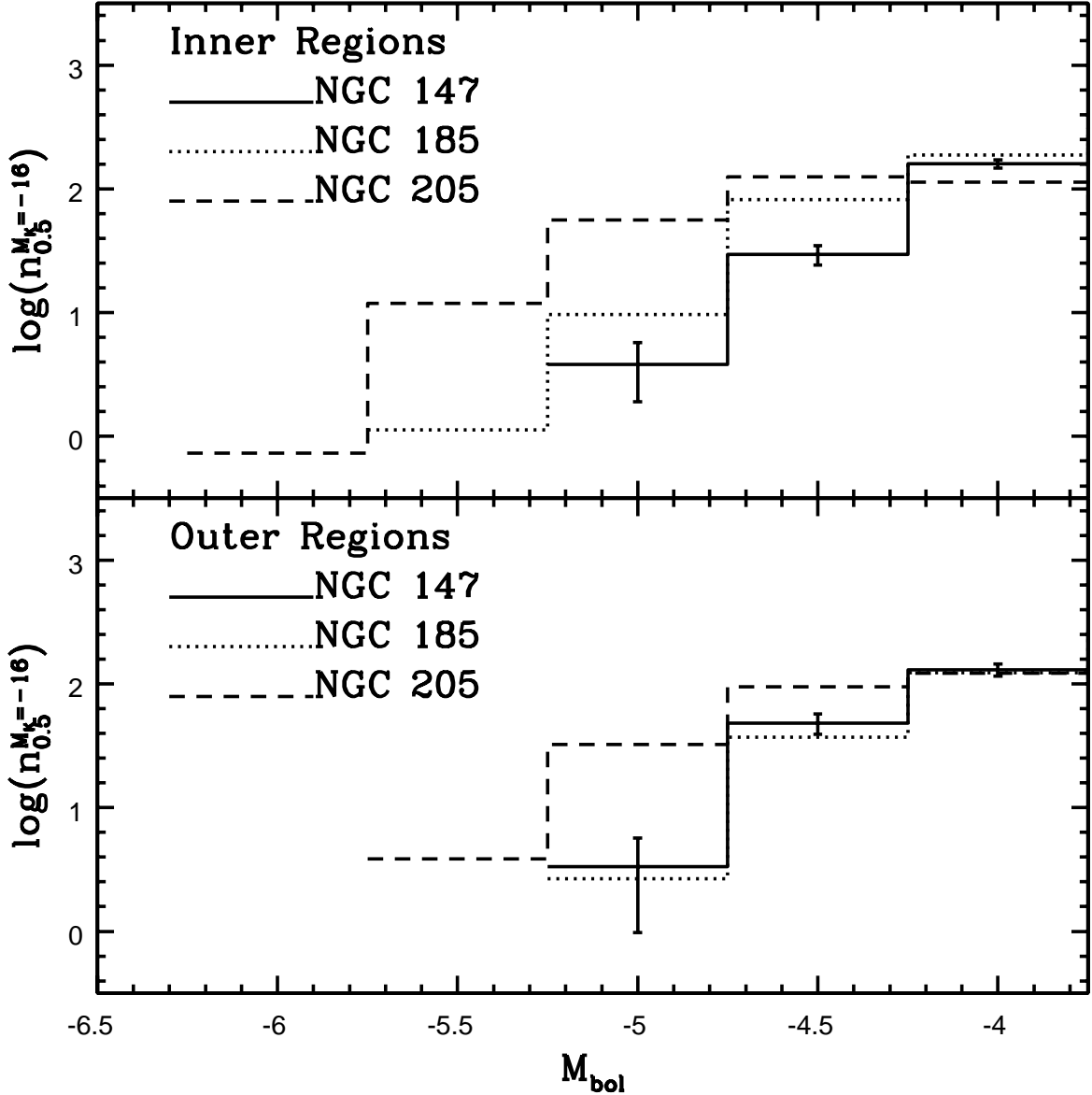}
\caption
{The specific frequency of M giants in NGC 147, NGC 185, and NGC 205, where 
$n_{0.5}^{M_K=-16}$ is the number of stars per 0.5 bolometric magnitude 
interval expected for a population with M$_K = -16$. 
Bolometric corrections were computed using the calibration for oxygen-rich stars 
in the Milky-Way and the LMC given by Bessell \& Wood (1984). The errorbars show 
the $1-\sigma$ Poisson uncertainties in the NGC 147 measurements.  Note that (1) 
the LFs of all three systems are in reasonable agreement near the faint end, where the 
M giant population contains stars that formed over a longer period of time than at the 
bright end of the LF, (2) NGC 205 has a much larger density of M giants with M$_{bol} < -4$ 
than either NGC 147 and NGC 185, and (3) the LFs of the outer regions of NGC 147 and NGC 185 
are in good agreement.}
\end{figure}

\clearpage

\begin{figure}
\figurenum{14}
\epsscale{1.0}
\plotone{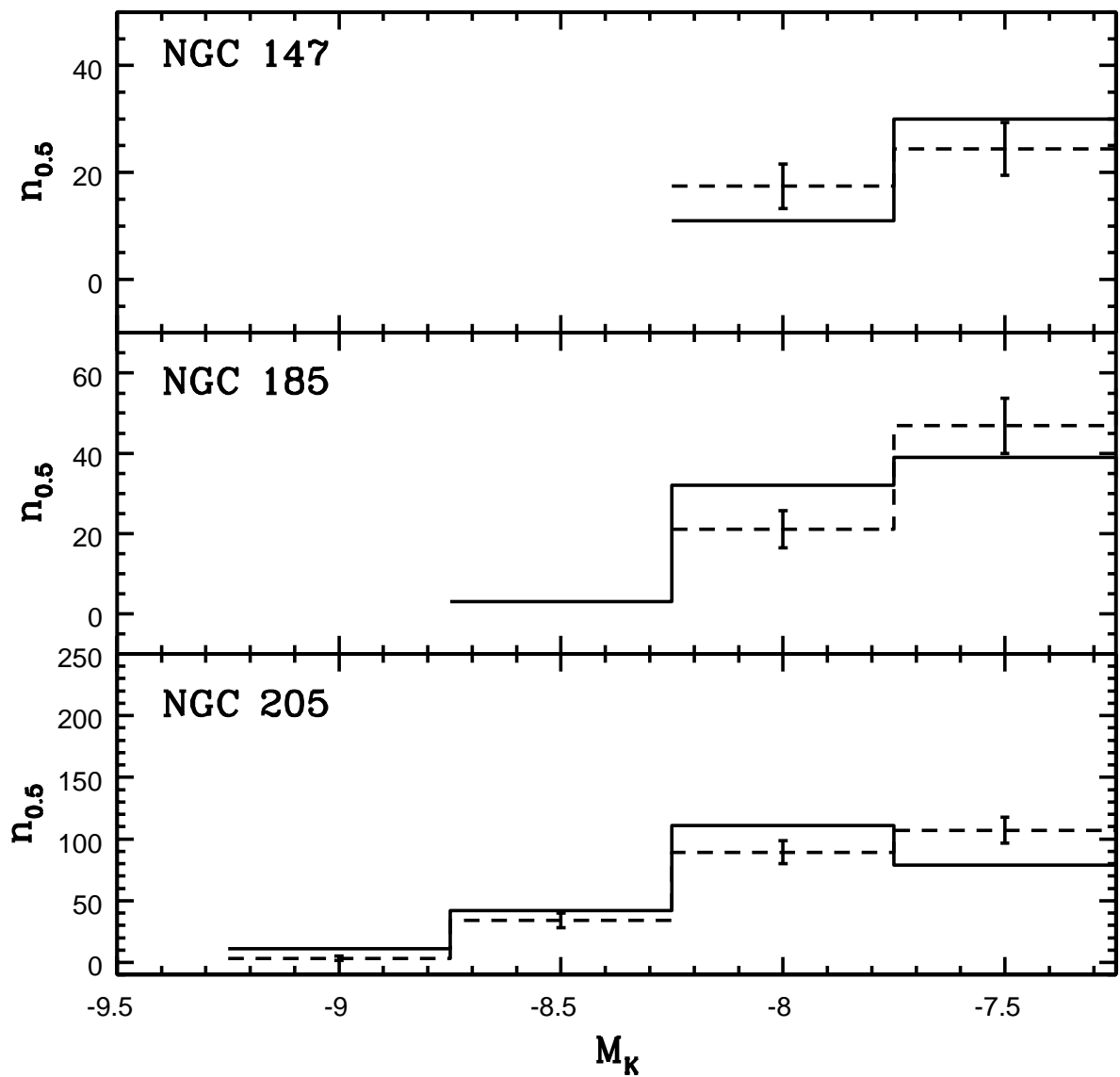}
\caption
{The M$_K$ LFs of C stars in the inner (solid lines) and outer 
(dashed lines) regions of NGC 147, NGC 185, and NGC 205; $n_{0.5}$ is the number 
of stars per 0.5 $K-$band magnitude interval in the inner region of each galaxy, and the 
errorbars show the $1-\sigma$ Poisson uncertainties in the outer region measurements. 
The number counts for the outer region data have been scaled to match 
the integrated $K-$band brightness in the inner region based on the $K-$band 
surface brightness profile in the 2MASS Extended Source Catalogue (Jarrett et al. 2000).}
\end{figure}

\clearpage

\begin{figure}
\figurenum{15}
\epsscale{1.0}
\plotone{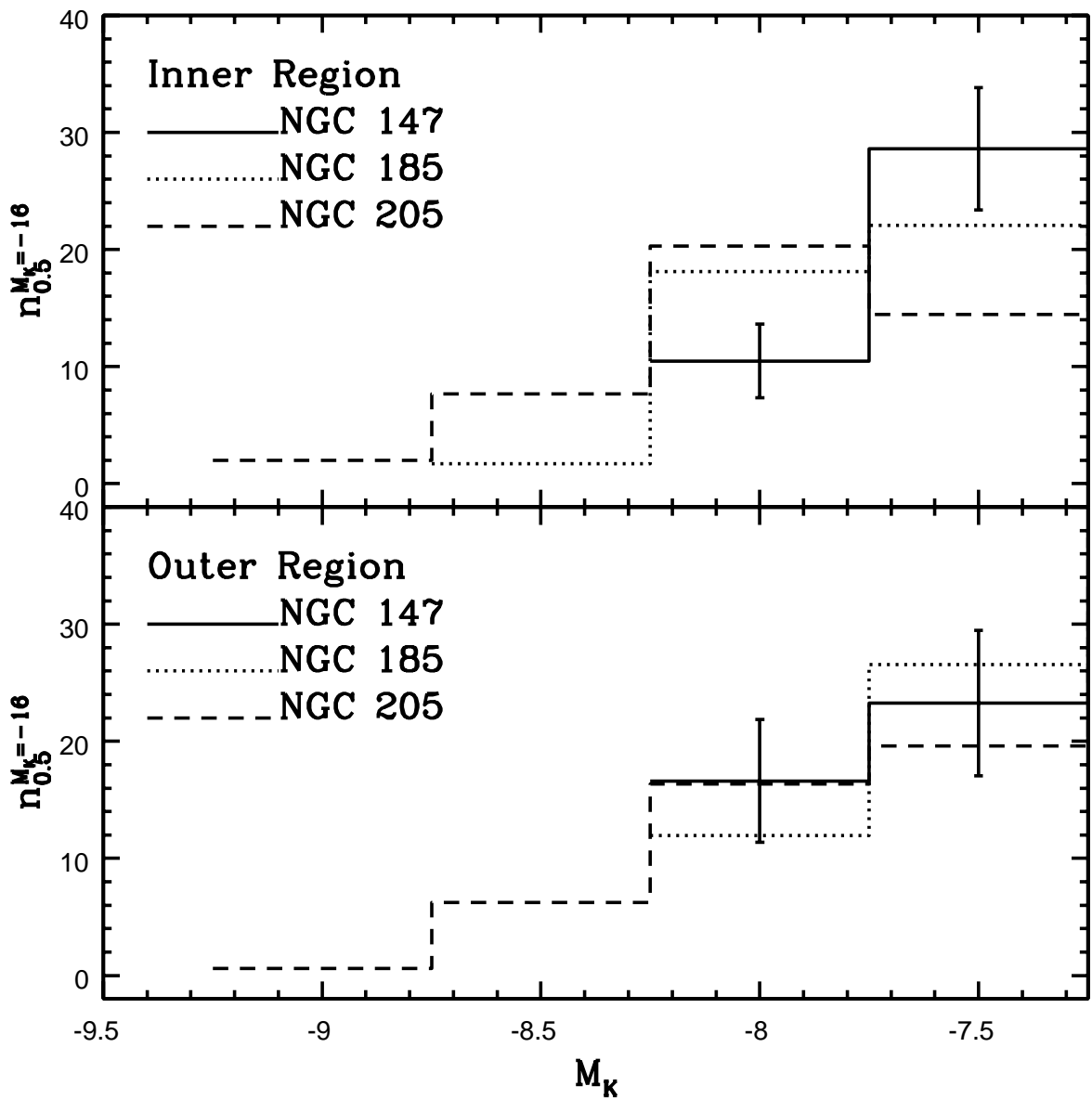}
\caption
{The specific frequency of C stars in NGC 147, NGC 185, and NGC 205 in the 
$K-$band; $n_{0.5}^{M_K=-16}$ is the number of stars per 0.5 
$K-$band magnitude interval expected for a population with M$_K = -16$, and the errorbars 
show the $1-\sigma$ Poisson uncertainties in the NGC 147 measurements. Note that (1) 
NGC 205 has a much higher density of C stars with M$_{K} < -8$ than either NGC 147 
and NGC 185, and (2) the specific frequency of C stars per M$_K$ bin in the inner and 
outer regions of NGC 147 and NGC 185 roughly agree within the estimated uncertainties.}
\end{figure}

\clearpage

\begin{figure}
\figurenum{16}
\epsscale{1.0}
\plotone{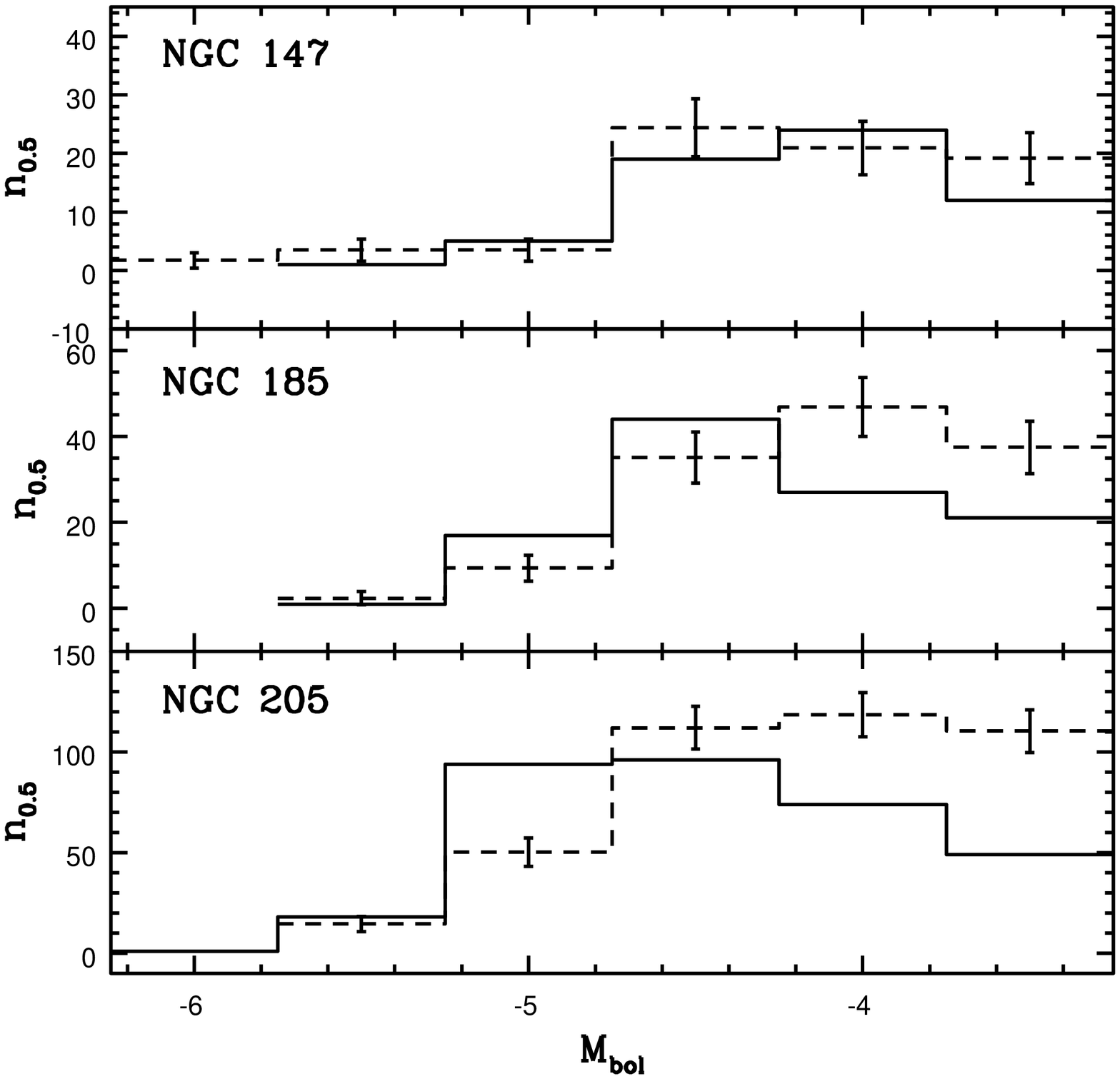}
\caption
{The M$_{bol}$ LFs of C stars in the inner (solid lines) and outer (dashed lines) regions 
of NGC 147, NGC 185, and NGC 205; $n_{0.5}$ is the number of stars per 0.5 bolometric 
magnitude interval, and the errorbars show the $1-\sigma$ Poisson uncertainties in the outer 
region measurements. The number counts for the outer region data have been scaled to match 
the integrated $K-$band brightness in the inner region based on the $K-$band 
surface brightness profile in the 2MASS Extended Source Catalogue (Jarrett et al. 2000).}
\end{figure}

\clearpage

\begin{figure}
\figurenum{17}
\epsscale{1.0}
\plotone{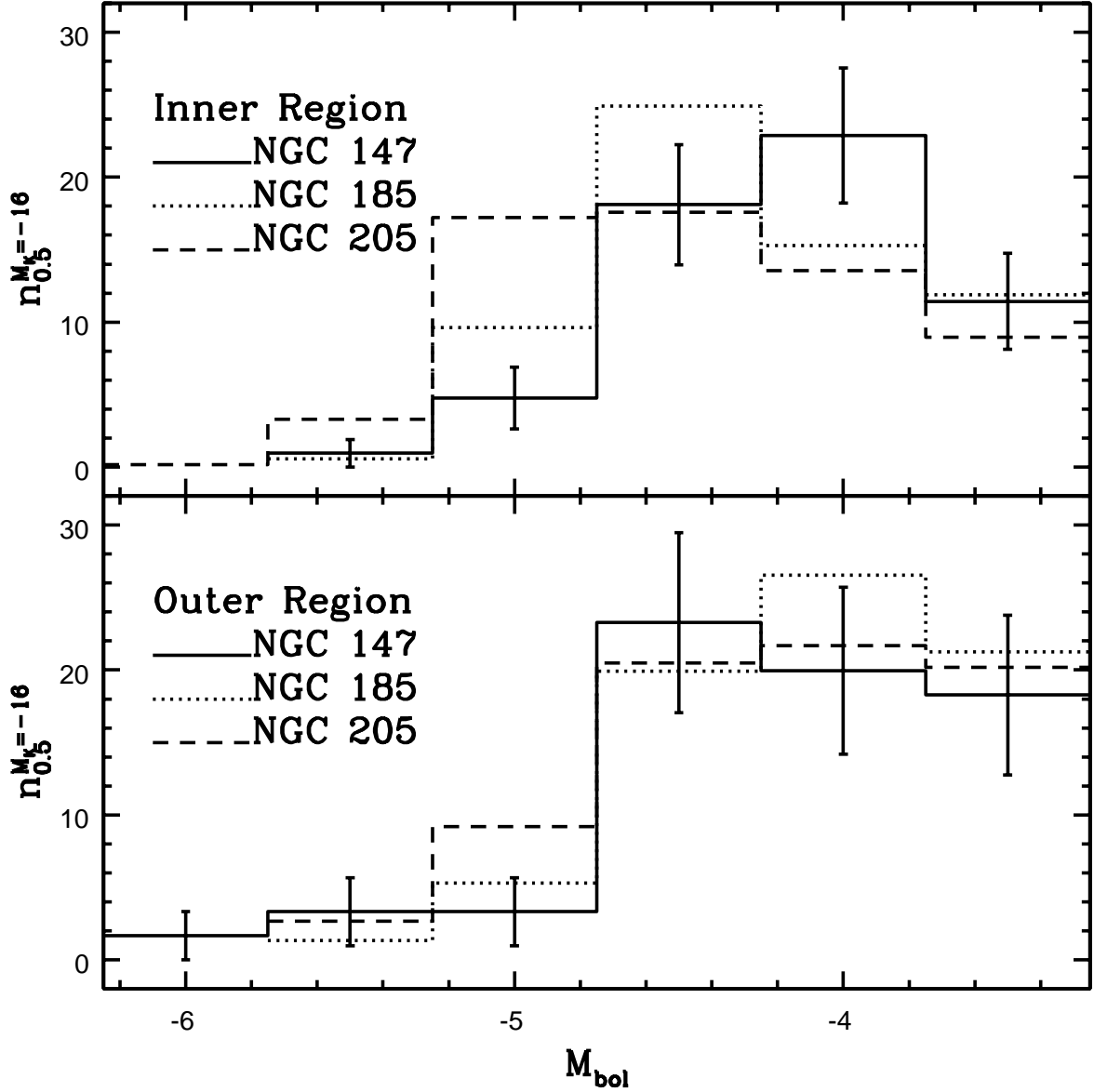}
\caption
{The M$_{bol}$ LFs of C stars in NGC 147, NGC 185, and NGC 205, scaled to 
match the number of objects in a source with an integrated brightness M$_K = -16$.
$n_{0.5}^{M_K=-16}$ is the number of stars per 0.5 bolometric magnitude interval expected for 
a population with M$_K = -16$, and the errorbars show the $1-\sigma$ Poisson 
uncertainties in the NGC 147 measurements. Note that 
while there are galaxy-to-galaxy differences among the inner region LFs, 
there is much better agreement between the outer region LFs. In fact, 
the specific frequencies of C stars at a given M$_{bol}$ 
outside the areas of most recent star formation agree well within 
the estimated $2-\sigma$ uncertainties.}
\end{figure}

\clearpage

\begin{figure}
\figurenum{18}
\epsscale{1.0}
\plotone{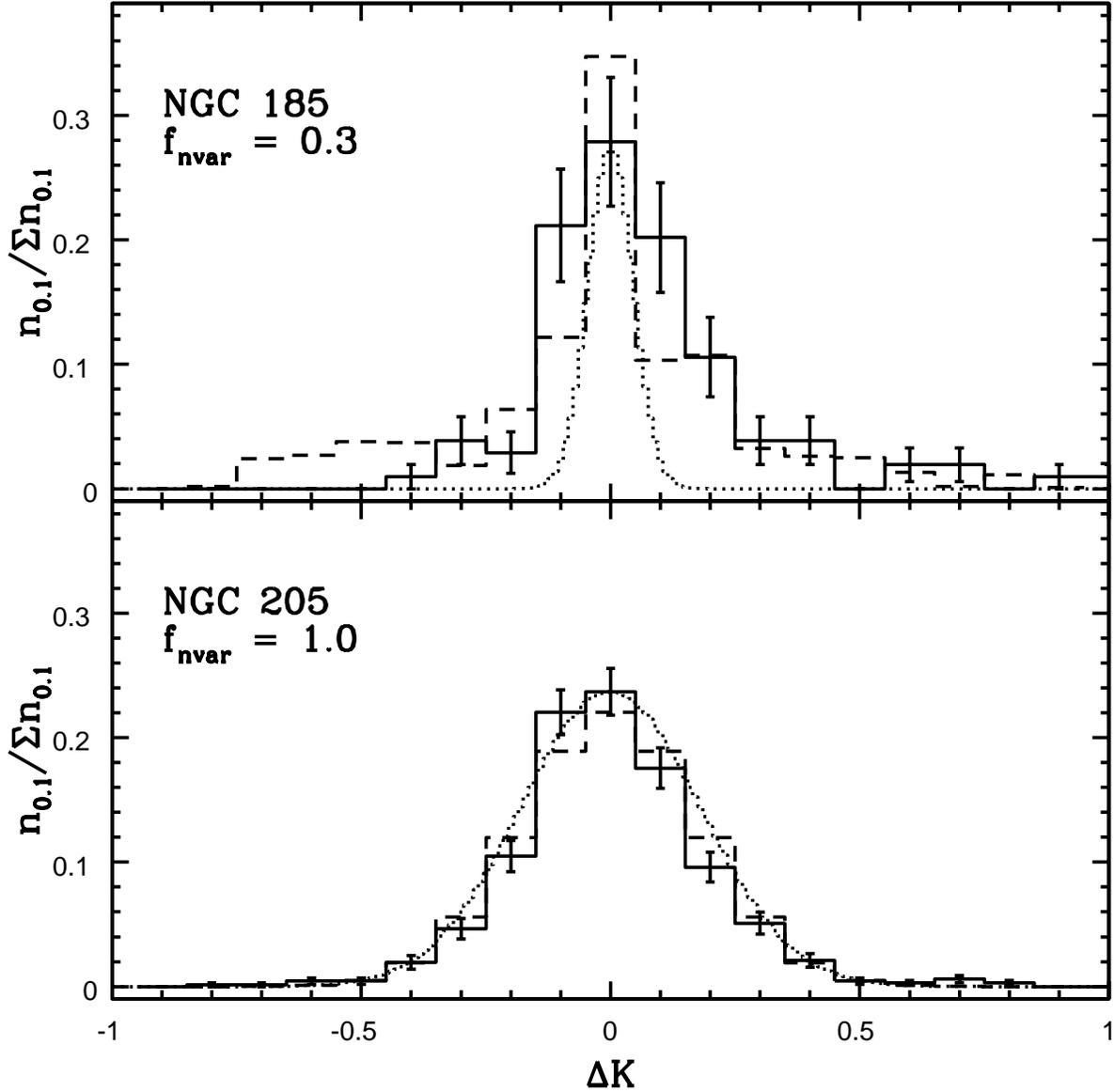}
\caption
{The histogram distributions of $\Delta K$ values for sources with M$_K$ between 
--7.5 and --8 in the inner regions of NGC 185 and NGC 205 (solid lines), 
compared with the predicted error distributions from artificial star experiments (dotted 
lines), which have been scaled to match the peak number of counts in the $\Delta K$ 
distribution. Also shown are the results of combining the $\Delta K$ values for LMC 
LPVs observed by Hughes \& Wood (1990) with a population of non-variable objects (dashed 
lines); the fractional size $f_{nvar}$ of the non-variable population that best 
matches the observations is given in the upper left hand corner of 
each panel. The model that best fits the NGC 185 data has $f_{nvar} = 0.3 \pm 0.15$ 
($1-\sigma$ uncertainty), and so is in good agreement with what is seen in other 
galaxies. While the model that best fits the NGC 205 data has $f_{nvar} = 1.0$, 
the scatter in the data is such that models with $f_{nvar} = 0.0$ 
can not be ruled out at the $2-\sigma$ level.}
\end{figure}

\end{document}